\documentclass[12pt,a4paper,DIV12]{scrartcl}
\usepackage[utf8x]{inputenc}
\usepackage[T1]{fontenc}
\usepackage{lmodern}
\usepackage[british]{babel}
\usepackage{graphicx}
\usepackage{grffile}
\usepackage{hyperref}
\usepackage{color}
\usepackage{slashed}
\usepackage{textcomp}
\usepackage{subfigure}
\usepackage{amsmath}
\usepackage{amssymb}
\usepackage{amsfonts}
\usepackage{lscape}
\usepackage{psfrag}

\DeclareOldFontCommand{\tt}{\normalfont\ttfamily}{\mathtt}
\DeclareOldFontCommand{\bf}{\normalfont\bfseries}{\mathbf}

\newcommand*{\email}[1]{\href{mailto:#1}{#1}} 
\newcommand{\arxiv}[1]{arXiv:\,\href{http://arxiv.org/abs/#1}{{\tt #1}}}
\newcommand{\betal}{\ensuremath{\beta_{\textrm{lat}}}}
\newcommand{\gl}{\ensuremath{g_{\textrm{lat}}}}
\newcommand{\ml}{\ensuremath{m_{\textrm{lat}}}}
\newcommand{\als}{\ensuremath{\alpha}}
      
\title{The running coupling from gluon and ghost propagators in the Landau gauge: Yang-Mills theories with adjoint fermions}
      
\author{%
Georg Bergner\\
\textit{\large Friedrich-Schiller-University Jena, Institute of Theoretical Physics}\\
\textit{\large Max-Wien-Platz 1, D-07743 Jena, Germany}\\
\textit{\large E-mail: \email{georg.bergner@uni-jena.de}}\\[5mm]
Stefano Piemonte\\
\textit{\large University of Regensburg, Institute for Theoretical Physics}\\
\textit{\large Universit\"atsstr.~31, D-93040 Regensburg, Germany}\\
\textit{\large E-mail: \email{stefano.piemonte@ur.de}}
\vspace*{5mm}}

\begin{document}

\maketitle

\begin{abstract}
Non-Abelian gauge theories with fermions transforming in the adjoint representation of the gauge group (AdjQCD) are a fundamental ingredient of many models that describe the physics beyond the Standard Model. Two relevant examples are $\mathcal{N}=1$ Supersymmetric Yang-Mills (SYM) theory and Minimal Walking Technicolor, which are gauge theories coupled to one adjoint Majorana and two adjoint Dirac fermions, respectively. While confinement is a property of $\mathcal{N}=1$ SYM, Minimal Walking Technicolor is expected to be infrared conformal. We study the propagators of ghost and gluon fields in the Landau gauge to compute the running coupling in the MiniMom scheme. We analyze several different ensembles of lattice Monte Carlo simulations for the SU(2) adjoint QCD with $N_f=1/2,\ 1,\ 3/2$, and 2 Dirac fermions. We show how the running of the coupling changes as the number of interacting fermions is increased towards the conformal window.
\end{abstract}

\section{Introduction}

Strong interactions are responsible for the confinement of quarks and for the generation of the hadron masses in QCD. The coupling ($g$ or $\als\equiv\frac{g}{4\pi}$) of the strong interactions as a function of the energy scale $\mu$ decreases at high energies and asymptotic freedom allows a description of the interactions in a perturbative framework. The coupling increases towards lower energies leading to color charge confinement, dynamical scale generation, and other non-perturbative effects. Asymptotic freedom is a general feature of many non-Abelian gauge theories. The determination of the running as a function of the scale, the gauge group, and the number of fermions is crucial for a deeper understanding of QCD and confinement. Moreover, it determines the applicability of possible extensions of Standard Model that solve for open issues related to unification of forces, the hierarchy problem, and dark matter. 

The running of the coupling with the scale $\mu$ is expressed by the $\beta$-function, that can be computed order by order in perturbation theory in a given scheme $\mathcal{S}$. This provides the asymptotic expansion
\begin{equation}\label{betaeq}
 \mu \frac{d}{d\mu} g^\mathcal{S}(\mu) = \beta^\mathcal{S}(g^\mathcal{S}(\mu)) = -\sum_{i=0}^{\infty} b_i^\mathcal{S} g^\mathcal{S}(\mu)^{3+i}\,.
\end{equation}
The first two coefficients $b_0$ and $b_1$ are scheme independent and asymptotic freedom requires $b_0$ to be positive. A truncation to the first few coefficients of the expansion provides a good approximation at high enough $\mu$, but it is not yet fully understood how the coupling of QCD runs in the deep infrared regime $\mu \rightarrow 0$. In particular many questions have been raised concerning the physical consequences of a singular behavior of $g^\mathcal{S}(\mu)$ at small scales in relation to confinement. Different possibilities have been discussed in literature, see for instance Ref.~\cite{DEU} for a review. The so-called infrared slavery scenario has been proved not to be strictly required for a realization of confinement.

Supersymmetric Yang-Mills theories (SYM) are a remarkable exception, since many non-perturbative features of these theories are known analytically. Instanton calculations lead to the conclusion that there exists a scheme where the $\beta$-function can be computed exactly to all-orders in perturbation theory. A well-known example is the NSVZ $\beta$-function \cite{NOV}
\begin{equation}
\label{eq:nsvzbeta}
\beta(g(\mu)) = - \frac{g(\mu)^3}{16 \pi^2}\frac{3 N_c}{1-\frac{N_c g(\mu)^2}{8 \pi^2}}\,,
\end{equation}
that provides the running of the strong coupling of $\mathcal{N}=1$ SYM. This theory is the minimal supersymmetric extension of the pure gauge sector of QCD describing the interactions between gluons and their supersymmetric partners, the gluinos, both transforming in the adjoint representation of the gauge group SU($N_c$). The running coupling of the NSVZ $\beta$-function includes all non-perturbative contributions due to the non-renormalization theorem \cite{ARK}.

The number of fermions $N_f$ has a negative contribution to the first two coefficients of the $\beta$-function $b_0$ and $b_1$, and there exists a critical number of fermions where $b_0$ is positive while $b_1$ is negative \cite{CAS,BAN}. In this case the running of the coupling freezes at some scale $\mu$ and the corresponding gauge theory is infrared conformal. It has been suggested that a near infrared conformal theory might be responsible for the breaking of electroweak symmetry \cite{DIM,EIC}. The Higgs boson would be a composite bound state, interpreted as a dilaton of the corresponding low energy effective theory. Many recent lattice calculations have been performed to understand how the conformal window is approached as the number of fermions $N_f$ is increased, exploring the phase space not only with fundamental quarks, but also with fermions in the adjoint and the sextet representation of the gauge group.

In this contribution we explore the non-perturbative dynamics of gluons and ghosts in the Landau gauge by means of numerical lattice simulations for an SU(2) gauge theory coupled to one, two, three and four Majorana fermions, corresponding to $N_f=1/2$, $3/2$, and $2$ Dirac fermions. We provide results for the running coupling in these theories defined in the Mini-MOM scheme. We investigate how the ghost and gluon propagators and the running coupling change when the number of fermions is increased starting from a confining theory, $\mathcal{N}=1$ SYM, to the point where the conformal window is reached at $N_f=2$.

There are several long-term goals of our investigations. The first one is to measure the running coupling of $\mathcal{N}=1$ SYM. Like in QCD, the running coupling and the scale $\Lambda$ is an important ingredient for further investigations, for example, of renormalized currents and condensates. In addition it might help to understand the implications of the exact results like \eqref{eq:nsvzbeta} for this theory.   

The second important goal is a general comparison of the running coupling in the MOM scheme for a QCD-like theory and a theory in the conformal window in order to see how the signs for an infrared fixed point appear in this setup. MWT is a prime example for such a comparison since the indications for a conformal behavior of this theory have been found in a large number of numerical investigations. So far this comparison of the running coupling in the MOM scheme has only been done using Dyson-Schwinger equations Ref.~\cite{HOP1,HOP2} for the case of $N_f$ fermions in the fundamental representation. We want to confront the scenario for the appearance of the infrared fixed point presented in these studies with our numerical results. 

The final purpose is to determine a scale $\Lambda$ from the running of the coupling at higher energies separated from the possible conformal scale invariant behavior in the infrared but also separated from the ultraviolet cutoff effects. This implies that we have to check whether we can connect the infrared and ultraviolet fixed points, i.~e.\ the nearly conformal and the perturbative running. We also have to consider the possibility of unphysical regions where a backward running of the coupling might be occur.

The third goal is an identification of the onset of the conformal window for theories with $N_f$ fermions in adjoint representation. The behavior of the running coupling indicates the appearance of a fixed point and provides hence a signal for the onset of the conformal window. The comparison of the running for different $N_f$ hence allows in principle to identify the critical $N_f$ for a transition from a confining to a conformal behavior.

This work is organized as follows. In the first part we start in the continuum and present the considered scheme for the running coupling in Sec.~\ref{sec:contrunning} and the considered theories in Sec.~\ref{sec:contadjqcd}. In Sec.~\ref{sec:runninglat} the procedure of our numerical determination of the running coupling on the lattice is explained. In the last sections  we present and discuss our non-perturbative results for the running coupling in the different theories, starting with the extreme cases $\mathcal{N}=1$ SYM (Sec.~\ref{sec:symres}) and MWT (Sec.~\ref{sec:nf2res}) before we consider the intermediate theories in Sec.~\ref{sec:nf32res} and \ref{sec:nf1res}.


\section{The running coupling in the MOM schemes and the solutions of Dyson-Schwinger equations}
\label{sec:contrunning}
In the continuum and in the Landau gauge, the form of gluon and ghost propagators, denoted in the following respectively as $G_{ab;\mu\nu}(p)$ and $D_{ab}(p)$, is constrained by Lorentz invariance. In particular, only two scalar functions $Z(p)$ and $J(p)$, called gluon and ghost dressing functions, are required to describe the deviations due to quantum interactions from the free non-interacting propagators. The renormalized gluon and ghost propagators have therefore the following structure:
\begin{eqnarray}
G_{ab;\mu\nu}(p) & = & \delta_{ab}\left(\delta_{\mu\nu} -\frac{p_\mu p_\nu }{p^2}\right) \frac{Z(p^2)}{p^2}\,,\\
D_{ab}(p) & = & \delta_{ab} \frac{J(p^2)}{p^2}\,.
\end{eqnarray}
In the Momentum (MOM) scheme, the renormalization condition for the ghost and the gluon dressing function is set such that the gluon and the ghost propagators are equal to their tree-level counterpart at the reference renormalization scale $\hat{\mu}^2 = p^2$. In other words, the renormalization condition is 
\begin{align}
\label{eq:rencond}
Z(\hat{\mu}^2) = J(\hat{\mu}^2) = 1\; .  
\end{align}

There are various possible alternative prescriptions for the renormalization of the coupling in the MOM schemes, which can be defined from the three or four gluon vertexes \cite{ALL,BOU2,BOU3,BOU4}, or from the quark-quark-gluon or ghost-ghost-gluon vertexes. In principle each definition of the coupling leads to a different running. The Landau gauge has the peculiar property that the renormalized coupling $\als$ at scale $\mu^2=p^2$ determined from the ghost-ghost-gluon vertex can be defined just in terms of the renormalized ghost and gluon dressing function as
\begin{equation}\label{ghmom}
\als(p^2) = \als(\hat{\mu}^2) Z(p^2) J(p^2)^2\,.
\end{equation}
The scheme is based on the UV-finiteness of the ghost-ghost-gluon vertex in the Landau gauge, accordingly to the so-called Taylor theorem \cite{TAY}. The main advantage of this scheme is that it requires the knowledge of only two-point functions, that can be reliably measured on the lattice with $O(20)$ gauge-fixed configurations.

The ``Taylor coupling'' defined by \eqref{ghmom} has been deeply investigated in QCD using non-perturbative lattice simulations and the Dyson-Schwinger approach \cite{SME1,BLC1,FIS1,BOU1}. The Landau gauge turns out to be advantageous for a numerical solution of the Dyson-Schwinger equations (DSEs). There has been a long discussion on whether the so-called ``scaling solution'' or ``decoupling solution'' of DSEs are effectively realized in confining non-Abelian gauge theories. The ``scaling solution'' predicts that the gluon and ghost dressing functions exhibit a power-like behavior in the infrared regime of the form
\begin{equation}\label{eq:scaling}
 Z(p) \sim (p^2)^{2\rho} ~~~~J(p) \sim (p^2)^{-\rho}\,,
\end{equation}
implying that the coupling runs to an infrared fixed point in the limit $p \rightarrow 0$ \cite{BLC1}. The ``decoupling solution'' implies that the gluon propagator has the form of Yukawa massive particles and pure Yang-Mills theory would be ``infrared trivial''. Lattice investigations have the tendency to prefer the ``decoupling solution'', although the infinite volume limit and the Gribov noise might be a limitation that prevents the observation of the ``scaling solution'' \cite{FIS2,BOG}.

The interactions of gauge fields with matter fields changes non-trivially the behavior of ghost and gluon propagators \cite{HOP1,HOP2}. There are two competing effects, namely the antiscreening of gluons and the screening of quarks. As the number of interacting fermionic degrees of freedom increases, the antiscreening of gluons still wins at high energies but it is not strong enough to overcome the screening provided by virtual quark loops at smaller energies. Therefore the running of the coupling freezes, and gluon and ghost propagators both diverge following a power law in the infrared limit. These conclusions arise from the solutions of coupled DSEs. In \cite{HOP1,HOP2} the onset of the conformal window for fermions in the fundamental representation has been investigated applying this Dyson-Schwinger approach. The result was a remarkably small critical $N_f=4.5$ for the onset of the conformal window. It was found that the dressing functions follow to a good approximation a polynomial form of type \eqref{eq:scaling} with $\rho=0.15$, which implies a decreasing $Z(p^2)$ and and increasing $J(p^2)$ towards the infrared. This has to be confronted with an alternative scenario in which both dressing functions become constant at the appearance of the infrared fixed point. We want to compare these scenarios with our numerical results. 

A four-loop perturbative calculation of the running of $\als(\mu)$ in the MiniMOM scheme has been presented for fermions in various representations including the adjoint one in Ref.~\cite{RYT}. The perturbative calculation will be used to compute the $\Lambda$-parameter of the adjoint QCD (AdjQCD) with different numbers of fermions in the MOM scheme and to verify whether our simulations are in a region where the lattice spacing is small enough such that the coupling at high energies follows the asymptotic scaling. We can disentangle the running of the bare lattice coupling $\gl$ as a function of the lattice spacing $a$ from the running of the strong coupling of \eqref{ghmom} as a function of the momentum, such that a clear separation of lattice cut-off effects from the ``physical'' continuum properties of the running coupling is possible.
Similar methods have been widely employed in previous lattice calculations of the running coupling for many BSM theories in the Schr\"odinger functional or the Wilson flow schemes \cite{ANN1,ANN2,ANN3,ANN4,ANN5,KUT1,KUT2,RAN}, however we will not attempt to use step scaling at directly zero fermion mass but we will compute the coupling from simulations at non-zero fermion mass produced by our collaboration to study the spectrum of bound states.

\section{The continuum action of adjoint QCD}
\label{sec:contadjqcd}
We study on the lattice gauge theories with fermions in the adjoint representation of the gauge group SU(2), which action reads in the continuum
\begin{equation}
S = \int d^4 x \left\{\frac{1}{4} (F_{\mu\nu}^a F_{\mu\nu}^a) + \frac{1}{2} \sum_{i=1}^{2 N_f}\bar{\lambda}_a^i \gamma^\mu D^{ab}_\mu \lambda_b^i\right\}\,.
\end{equation}
The field $\lambda$ fulfills the Majorana condition
\begin{equation}
\bar{\lambda}_a = \lambda^T_a C\,,
\end{equation}
and the covariant derivative acts in the adjoint representation as
\begin{equation}
 D^{ab}_\mu \lambda_b = \partial_\mu \lambda_a + i g A_\mu^c (T_c^A)^{ab} \lambda_b\,,
\end{equation}
where $T_c^A$ are the Lie algebra generators as given by the structure constants. In the following we use the Dirac counting of fermionic degrees of freedom, meaning that for instance $N_f=2$ corresponds to AdjQCD with two Dirac or four Majorana fermions. A sign problem can arise for odd number of fermions, but our simulations are performed in a range of fermion masses where the sign problem can be neglected.

The general properties of adjoint QCD depend strongly on the number of fermions. There are three relevant possibilities:

\begin{itemize}
 \item \textbf{$N_f=1/2$ adjoint QCD}
 
 $\mathcal{N}=1$ supersymmetric Yang-Mills theory has only a single Majorana fermion, the gluino, interacting with the gluon. The gluon and the gluino field are related by a supersymmetriy transformation, which assumes on-shell the form
\begin{eqnarray}\label{susytr}
A_\mu(x) & \rightarrow & A_\mu - 2 i \bar{\lambda}\gamma_\mu \epsilon\,, \\
\lambda_a & \rightarrow & \lambda_a - \sigma_{\mu\nu} F^{\mu\nu}_a \epsilon\,,
\end{eqnarray}
where $\sigma_{\mu\nu}$ is proportional to the commutator $[\gamma_\mu,\gamma_\nu]$ and $F_{\mu\nu}$ is the field strength. The infinitesimal SUSY transformation is parametrized by a global Majorana spinor $\epsilon$. Lattice calculations have shown that the theory is confining, in the sense that there is a linearly rising of the potential between two static fundamental quarks which defines a string tension $\sigma$. The bound state spectrum is organized in supermultiplets of particles with equal masses, and recently Monte Carlo simulations have been able to show that supersymmetry is intact also at non-perturbative level and restored on the lattice in the continuum limit $a \rightarrow 0$ \cite{BER1}. 

\item \textbf{$N_f=2$ adjoint QCD}

$N_f=2$ AdjQCD is also known as Minimal Walking Technicolor \cite{SAN}. Predicted to be near the conformal window, the theory would be a good candidate for electroweak symmetry breaking to the Standard Model without the Higgs field, while satisfying at the same time the experimental constraint on the flavor-changing neutral currents. There have been several studies of the properties of the model, a conformal or near conformal behavior has been observed in particular from the scaling of the masses of bound states \cite{DEB1,HIE,BER2}. The glueball $0^{++}$ has been found to be the lightest bound state, however the observed mass anomalous dimension $\gamma^* \approx 0.3$ would be too small for realistic particle phenomenology. The determination of the running coupling using step-scaling and the Schr\"odinger functional method has provided evidences for the existence of an infrared fixed point of the $\beta$-function \cite{DEG,RAN}.

\item \textbf{$N_f=3/2$ and $N_f=1$ adjoint QCD}

Very little is known about the properties of $N_f=1$ AdjQCD. It can be considered as the limit of $\mathcal{N}=2$ SYM in which the scalars get an infinite mass and decouple. $\mathcal{N}=2$ SYM is known to be confining, however the confinement properties could be spoiled if SUSY is broken. Some preliminary studies of $N_f=1$ AdjQCD on the lattice have been presented in Ref.~\cite{ATH}, observing a scaling of the bound spectrum compatible with a conformal or near conformal theory with a large mass anomalous dimension $\gamma^* \approx 1$. The large mass anomalous dimension would be useful for particle phenomenology, however the breaking patter would not provide the required Goldstone bosons for the breaking of electroweak symmetry, and further fermion flavors in the fundamental representation must be included. 

To the best of our knowledge, there has been no previous lattice investigations of AdjQCD with three Majorana fermions. The theory is predicted to be just close to the conformal window. The main interest for $N_f= 3/2$ is therefore to understand how the conformal properties of adjoint theories depends on the number of flavors.

\end{itemize}

\section{Gauge fixing and correlators in momentum space on the lattice}
\label{sec:runninglat}
Our numerical simulations are done on a four-dimensional lattice of $(L_x, L_y, L_z, L_t)$ points in each direction. We take the same lattice size $L_s$ for the spacial directions and a larger temporal extend $L_t$. The bare gauge coupling of the lattice theory is $\gl$ and the inverse bare coupling is defined as $\betal=\frac{2N_c}{\gl^2}$. The bare fermion mass $\ml$ is fixed by the Hopping parameter $\kappa=\frac{1}{4\ml+8}$.

On the lattice, the Landau gauge is fixed by maximizing the functional
\begin{equation}
 \Pi(\{U\}) = \sum_{x,\mu} \textrm{Re} (\textrm{Tr}(\Omega(x) U_\mu(x) \Omega^\dag(x+\mu)))\,,
\end{equation}
with respect to the gauge transformation $\Omega(x)$. The gluon fields $A_\mu(x)$ are related to the traceless-antihermitian (TA) part of the gauge links $U_\mu(x)$ by the definition
\begin{equation}
 A_\mu(x) = \left.\frac{1}{i \gl} U_\mu(x)\right|_{\textrm{TA}}\,.
\end{equation}
We employ a sequence of standard overrelaxation updates to find the maximal of the functional $\Pi(\{U\})$ \cite{MAN}, together with parallel tempering. We stop when the condition
\begin{equation}
 \sum_{x,\mu} (\Delta_\mu A_\mu)^2 < 10^{-14}
\end{equation}
is satisfied. Many different $\Omega(x)$ configurations will satisfy the above condition on the lattice and there has been several studies to understand the impact of this uncertainty. Some small effect induced by the Gribov noise has been discovered in the deep infrared region of QCD for large volumes in Ref.~\cite{STE1}, however the full interpretation of the gauge fixing effects on the lattice has been debated, see also Ref.~\cite{MAS1}.

Once the gauge-fixed links are known, the gluon propagator in the momentum space can be easily computed from the Fourier transform of the gluon field in the real space. The computation of the ghost propagator requires an inversion of the Faddeev-Popov operator for each momentum of the ghost field. We solve the corresponding linear system using the BiCGStab biconjugate gradient algorithm. The definition of the Faddeev-Popov operator on the lattice is discussed in detail in Ref.~\cite{STE1}.

We compute the gluon and ghost propagators for each ensemble on $O(20)$ gauge-fixed configurations well separated by at least fifty molecular dynamics units. As described in \cite{SME1,BLC1}, the running coupling can be computed directly just in terms of the lattice bare ghost and gluon dressing function in the MiniMOM scheme as
\begin{equation}\label{minimom}
\als p^2 = \frac{\gl^2}{4 \pi} Z_0(p^2) J_0(p^2)^2\,.
\end{equation}
The running scale is defined by the lattice momentum and the renomalization scale is set at the inverse lattice spacing. 

\begin{table}
\centering
\begin{tabular}{ c | c | c | c | c | c}
  $N_f$ & $\betal$ & $\kappa$ & $L_s^3\times L_t$            & $am_{\textrm{PCAC}}$ & $am_\pi$ \\ \hline
  1/2   & 1.75    & 0.1492   & $32^3\times64$ & --                   & 0.20275(74) \\
  1/2   & 1.9     & 0.14387  & $32^3\times64$ & --                   & 0.21410(33) \\
  1/2   & 1.9     & 0.14415  & $32^3\times64$ & --                   & 0.17520(22) \\
  1/2   & 1.9     & 0.14435  & $32^3\times64$ & --                   & 0.14129(59) \\ \hline
  1     & 1.75    & 0.1650   & $16^3\times32$ & --                   & --          \\
  1     & 1.75    & 0.1650   & $24^3\times48$ & 0.062837(79)         & 0.4648(12)  \\
  1     & 1.75    & 0.1660   & $24^3\times48$ & 0.03567(11)          & 0.3313(11)  \\ \hline
  3/2   & 1.5     & 0.1330   & $24^3\times48$ & 0.19515(20)          & 0.86625(73) \\
  3/2   & 1.5     & 0.1340   & $24^3\times48$ & 0.15632(15)          & 0.74286(62) \\
  3/2   & 1.5     & 0.1351   & $24^3\times48$ & 0.10986(12)          & 0.58219(99) \\
  3/2   & 1.6     & 0.1300   & $24^3\times48$ & 0.19161(26)          & 0.7966(15)  \\
  3/2   & 1.7     & 0.1285   & $32^3\times64$ & 0.173655(41)         & 0.69268(25) \\
  3/2   & 1.7     & 0.1300   & $32^3\times64$ & 0.129098(37)         & 0.55712(19) \\
  3/2   & 1.7     & 0.1320   & $32^3\times64$ & 0.06635(12)          & 0.3312(20)  \\ \hline
  2     & 1.5     & 0.1325   & $32^3\times64$ & 0.128840(55)         & 0.58848(98) \\ 
  2     & 1.5     & 0.1335   & $32^3\times64$ & 0.089619(74)         & 0.44212(28) \\ 
  2     & 1.5     & 0.1350   & $32^3\times64$ & 0.030414(45)         & 0.17063(65) \\ 
  2     & 1.7     & 0.1275   & $32^3\times64$ & 0.17697(22)          & 0.66093(22) \\ 
  2     & 1.7     & 0.1285   & $32^3\times64$ & 0.147091(22)         & 0.57247(16) \\ 
  2     & 1.7     & 0.1300   & $32^3\times64$ & 0.100878(47)         & 0.42116(32) \\ 
  2     & 1.7     & 0.1300   & $24^3\times64$ & 0.101728(79)         & 0.4039(86)  \\  
  2     & 2.25    & 0.1250   & $32^3\times64$ & --                   & --          \\
  2     & 2.25    & 0.1275   & $32^3\times64$ & --                   & --          \\
\end{tabular}
\caption{Summary table of the analyzed ensembles. In the case of $N_f=1/2$, the PCAC mass is not measured and the pion mass is defined in a partially quenched approach \cite{MUN}.
The lattice action consists of a tree-level Symanzik improved gauge action and Wilson fermions in the adjoint representation. The simulations of $N_f=1/2$ are with one level, at $N_f=2$ and $3/2$ with three level of stout smearing.
The simulations of $N_f=1$ are done with tree-level clover improvement.
}\label{TABSUM}
\end{table}

There are several possible equivalent definitions of the modulus squared of the lattice momentum $\hat{p}$ 
\begin{equation}
 \hat{p} = 2\pi \left\{\frac{n_x}{L_x}, \frac{n_y}{L_y}, \frac{n_z}{L_z}, \frac{n_t}{L_t}\right\}\,,~~~~n_i \in (-L_i/2, L_i/2]\,.
\end{equation}
In the following we always plot scales of lattice momenta defined as 
\begin{equation}
 a^2 p^2 = \sum_{i=0}^3 \left\{2\sin\left(\frac{a p_i}{2}\right)\right\}^2\,.
\end{equation}
Lorentz invariance is broken on the lattice and therefore results coming from momenta with the same $a^2 \mu^2$ but different lattice directions will in general differ. The leading terms that violate Lorentz invariance are proportional to
\begin{equation}
 \sum_\mu (a p_\mu)^4\,,
\end{equation}
and it can be easily proven that for a given $\mu^2$, the above expression is minimized if the momentum is equally distributed on all four components, i.e. if the most symmetric momenta are chosen. There are several different criteria to cut the on-axis momenta. A possibility is to perform a cylindric cut to the momenta we measure, excluding directions where for instance $(p \cdot d / \sqrt{p \cdot p}) < 0.95$, with $d = \{1/2,1/2,1/2,1/2\}$. We choose instead to constrain the maximal absolute value of the deviation from the diagonal direction, defined as
\begin{equation}
\sum_\mu \left|\hat{p}_\mu - \frac{1}{4}\sum_\nu \hat{p}_\nu \right| < 0.55\,.
\end{equation}
Compared to the cylindric cut, the condition above allows more off-diagonal directions in the low momentum region, that are still interesting to understand the general behavior at low energy of gluon and ghost propagators. Field propagators at low momenta are usually affected more from finite volume effects rather than from lattice artefacts. Finally, we exclude the lattice modes with $a^2 p^2$ larger than four, to avoid ultraviolet regions affected by strong lattice artefacts.

The running of the strong coupling is given by the $\beta$-function according to the relation
\begin{equation}
 \frac{\mu}{\Lambda} = \left(\frac{b_0g(\mu)^2 }{16 \pi^2} \right)^{\frac{b_1}{2b_0^2}} \exp{\left(\frac{16 \pi^2}{2 b_0 g(\mu)^2}\right)} \exp{\left\{\int_0^{g(\mu)} \left(\frac{1}{\beta(g')} + \frac{16 \pi^2}{b_0 g'^3} - \frac{b_1}{b_0^2 g'} \right) d g'\right\}}\,.
\end{equation}
The parameter $\Lambda$ appears as an integration constant of the differential of \eqref{betaeq} and it depends on the scheme. The non-perturbative estimation of $4 \pi \als(\mu) = g(\mu)^2$ allows to determine the $\Lambda$-parameter, assuming a sufficiently high energy where the running given by the perturbative $\beta$-function holds. The so-called ``window problem'' is related to the difficulties of finding in current lattice calculations such an high energy region where $\mu$ is at the same time still far away from the cut-off $1/a$ where lattice artefacts are dominant. 

The fit to perturbation theory is performed including the functional form of the perturbative lattice correction \cite{STE3, SOT}
\begin{equation}
 \als(ap_\mu, p^2) = \als^{\textrm{pert}}(p^2)\left\{1 + \left(k_1 + \frac{k_2}{p^2} \right)\sum_\mu p_\mu^4 + k_3 \sum_\mu p_\mu^6\right\}\,.
\end{equation}
The correction to take into account the breaking of Lorentz symmetry on the lattice. This functional form stabilizes and reduces the $\chi^2/\textrm{d.o.f.}$ of the fit. The constants $k_n$ are determined from a global fit of the running coupling and are equal for all lattice momenta. We check the consistency of the estimated $\Lambda$-parameter by performing the fit to pertubation theory without lattice corrections, considering only the lattice momenta in strictly diagonal directions. In addition, in the MiniMOM scheme, the renormalization constant of the ghost-ghost-gluon vertex is set to one, therefore we must also fit the absolute normalization of $\als(\mu)$ together with $\Lambda$.

We employ in all our simulations for all theories the tree-level Symanzik improved gauge action and stout-smeared Wilson fermions, except for the $N_f=1$ AdjQCD where we use tree-level clover improved Wilson fermions. The simulations of $N_f=1/2$ are with one level, at $N_f=2$ and $3/2$ with three level of stout smearing. The summary of the ensembles analyzed is presented in Table~\ref{TABSUM}. While free gluon propagators have already an $O(a^4)$ improved form, there are lattice artefacts of the order $O(\gl^2 a^2)$ appear if one leaves the weak coupling regime. In addition, the interaction of gauge fields with Wilson fermions will introduce lattice artefacts of the order of $O(a)$ or $O(\gl^2 a)$ in the extrapolation to the continuum limit of the renormalized gluon and ghost propagators.

\section{$\mathcal{N}=1$ supersymmetric Yang-Mills theory}
\label{sec:symres}
The study of $\mathcal{N}=1$ SYM has been the subject of a long project of the DESY-M\"unster collaboration. The early simulations at $\betal=1.6$ failed to observe the formation of supermultiplets and the expected degeneracy in the particle spectrum due to a too coarse lattice spacing \cite{DEM}. Restoration of supersymmetry on the lattice has been recently reported in the continuum limit including results at $\betal=1.75$ and $\betal=1.9$ \cite{BER1,BER3,BER4}. Therefore we are going to focus the analysis of the gluon and ghost propagators only on these last two sets of ensembles, excluding the lattices generated at $\betal=1.6$.

\begin{figure}
\centering
 \subfigure[$\betal=1.9$ and $V=32^3\times 64$]{\includegraphics[width=.47\textwidth]{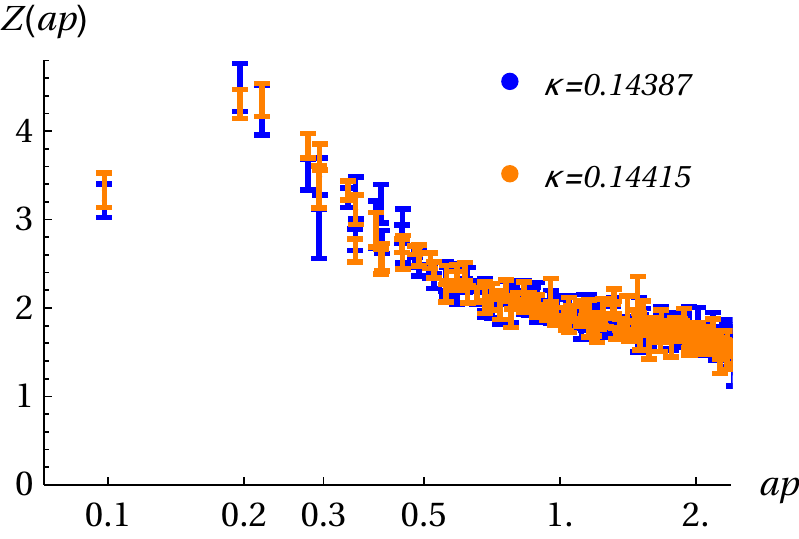}\label{symgluon190}}
 \subfigure[$\betal=1.9$ and $V=32^3\times 64$]{\includegraphics[width=.47\textwidth]{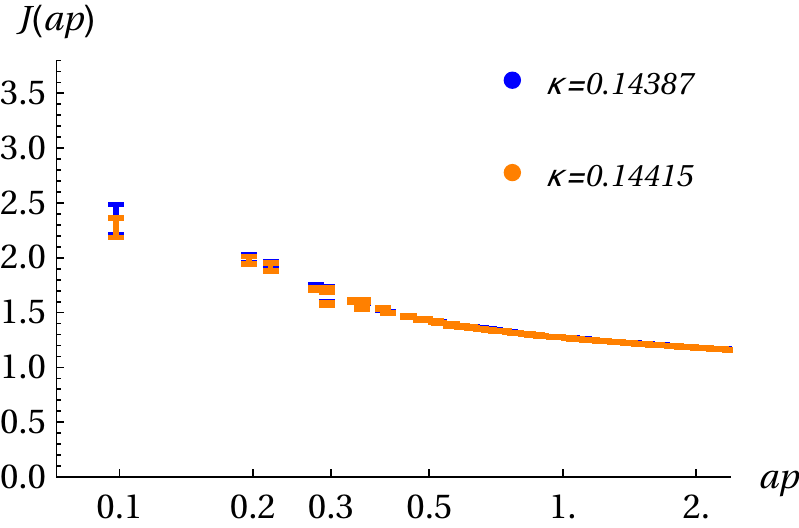}\label{symghost190}}
 \subfigure[$V=32^3\times 64$]{\includegraphics[width=.47\textwidth]{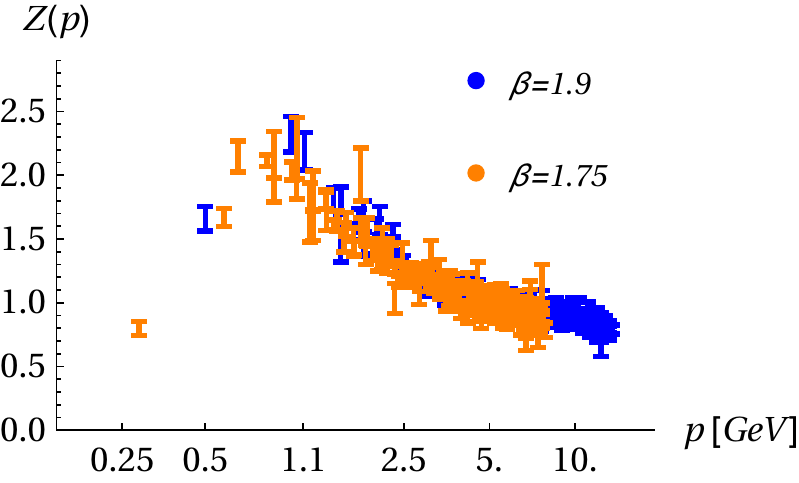}\label{symgluonreno}}
 \subfigure[$V=32^3\times 64$]{\includegraphics[width=.47\textwidth]{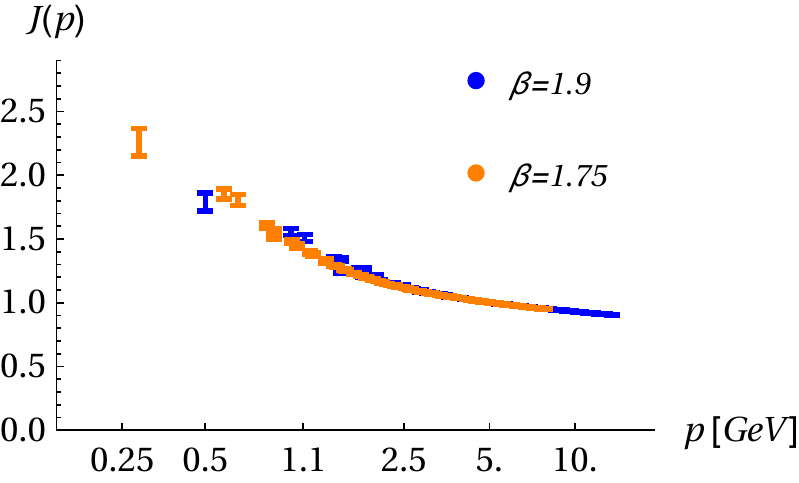}\label{symghostreno}}
 \caption{a-b) Bare gluon and ghost dressing functions for $\mathcal{N}=1$ SYM as a function of the lattice momentum $ap$ at $\betal = 1.9$. c-d) Renormalized gluon and ghost dressing functions for $\mathcal{N}=1$ SYM  on a logarithmic scale of the physical momentum $p$. The gluon dressing function tends to zero for $p\rightarrow 0$.}
\end{figure}
The bare gluon and the ghost dressing functions are shown at fixed $\betal=1.9$ in Fig.~\ref{symgluon190} and Fig.~\ref{symghost190}. There is no dependence of $J(ap)$ and $Z(ap)$ on the gluino mass up to our statistical accuracy. The renormalized gluon and ghost dressing functions are shown in Fig.~\ref{symgluonreno} and Ref.~\ref{symghostreno}. The renormalization condition \eqref{eq:rencond} is set at $\mu = 5.1$ GeV in QCD units, which means setting at that scale bare quantities equal to their tree-level counterpart, i.e. equal to one in this case. We use $w_0(\betal=1.9)/a = 5.858(84)$, $w_0(\betal=1.75)/a = 3.411(18)$ and $w_0 = 0.226$ fm to set the scale \cite{BER5}. The conversion of the momenta in physical units is effectively a different rescaling of the $x$-axis for each $\betal$, while the renormalization condition requires effectively a rescaling of the $y$-axis. Our results coming from the two different bare lattice couplings fall onto a unique curve, proving the multiplicative renormalization of $Z(p)$ and of $J(p)$. In particular, $Z(p)$ has a downward tendency at the smallest $\mu$ accessible with our volumes at $\betal=1.75$. The ghost propagator appears instead to be divergent in the deep infrared, and the ghost dressing function has an upward tendency in the same regime.

\begin{figure}
\centering
 \subfigure[$\betal=1.9$ and $V=32^3\times 64$]{\includegraphics[width=.47\textwidth]{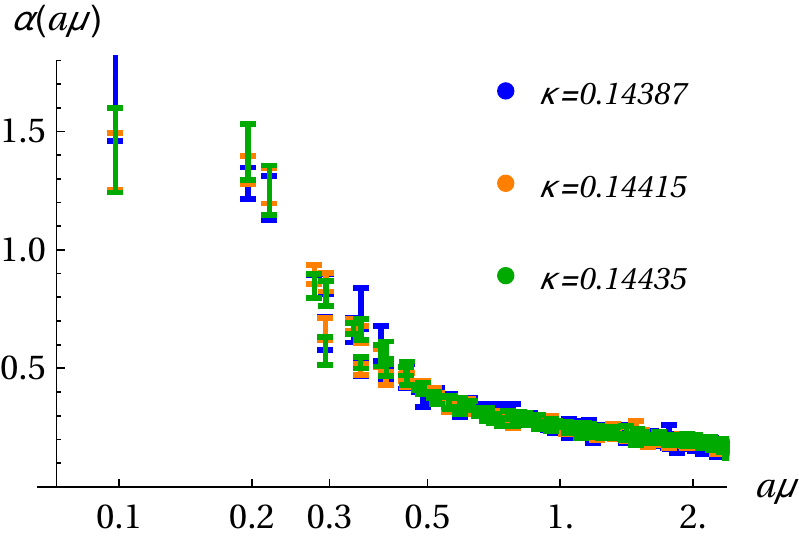}\label{alphasym1900}}
 \subfigure[$\betal=1.75$ and $V=32^3\times 64$]{\includegraphics[width=.47\textwidth]{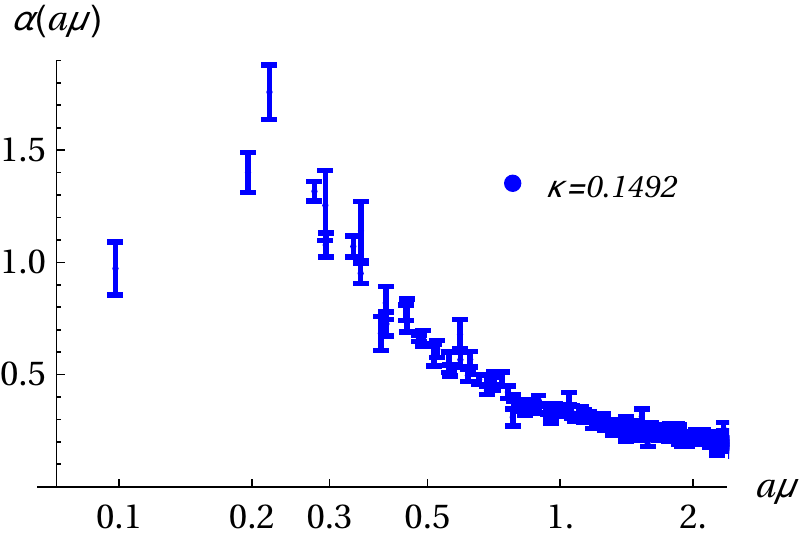}}
 \caption{a-b) The running of the running coupling $\als(\mu)$ for the $\mathcal{N}=1$ SYM theory. There is only a very weak dependence of $\als(\mu)$ on the gluino mass.
 }\label{symalpha}
\end{figure}
The running coupling defined in \eqref{minimom} is shown in Fig.~\ref{symalpha}. It decreases at high energy as predicted by asymptotic freedom, but, as observed in QCD, at very small $\mu$ the upward tendency of the ghost dressing function is not sufficient to compensate the downward tendency of the gluon dressing function, so that the running coupling $\als(\mu)$ peaks in the infrared at around $ap\simeq 0.2$, corresponding to $\mu\approx 1$ GeV in QCD units.
More extensive simulations at larger volumes are required to address the issue of the running coupling of the $\mathcal{N}=1$ SYM in the deep infrared.

\begin{figure}
\centering
 \subfigure[$\betal=1.9$ and $\kappa=0.14415$]{\includegraphics[width=.47\textwidth]{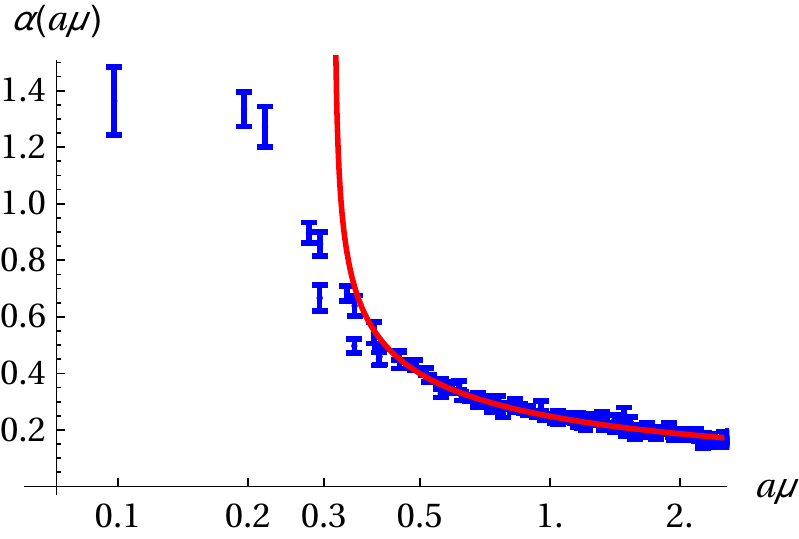}}
 \subfigure[$\betal=1.9$ and $\kappa=0.14415$]{\includegraphics[width=.47\textwidth]{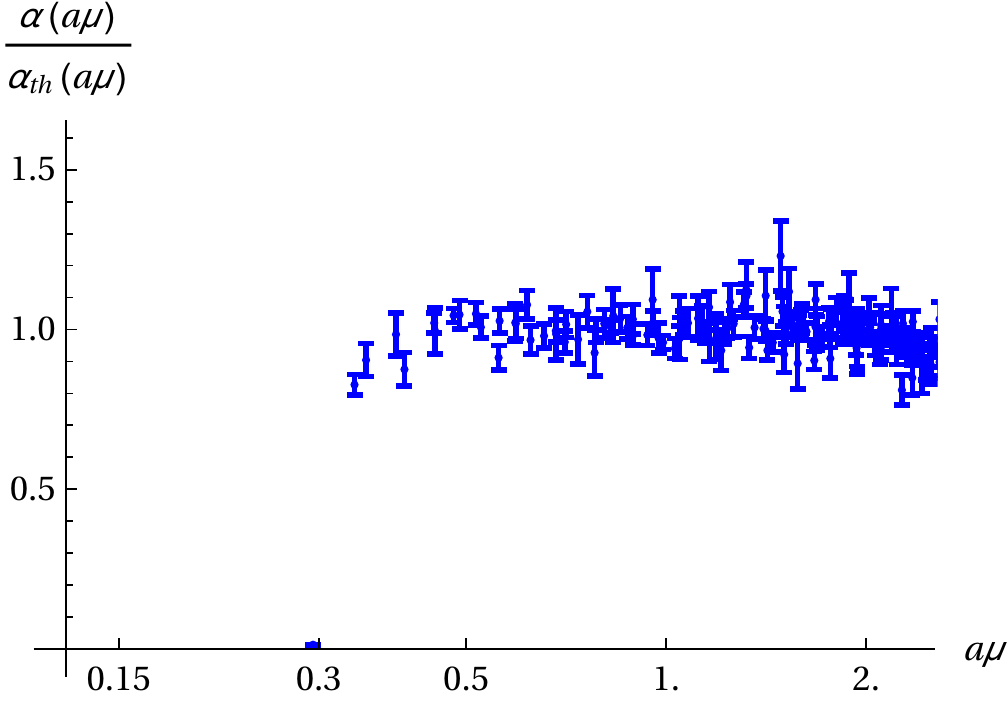}}
 \caption{The running of the running coupling $\als(\mu)$ for $\mathcal{N}=1$ SYM. The running can be related to four-loop perturbation theory at high energy. As shown on the left picture the ratio of the lattice results over the perturbative prediction is constant. Consequently there is a fairly large window to where perturbation theory can be trusted.}
\end{figure}
According to the methods presented in Sec.~\ref{sec:runninglat} we obtain for the scale $\Lambda$ the values
\begin{eqnarray}
a\Lambda^{\textrm{MiniMOM}}(\betal=1.75)&=&0.17(3)\,,\\
a\Lambda^{\textrm{MiniMOM}}(\betal=1.9)&=&0.120(15)\,,
\end{eqnarray}
where the quoted errors are purely systematic, coming from various choices of the fitting ranges. These scales are of the same order of magnitude as the mass of the lightest particle. When extrapolated to the chiral limit the lightest mass in lattice units is around $0.228(71)$ at $\betal=1.7$ and around $0.174(14)$ at $\betal=1.9$. 
The dimensionless $\Lambda$-parameter in the MiniMOM scheme in $w_0$ units,
\begin{eqnarray}
w_0\Lambda^{\textrm{MiniMOM}}(\betal=1.75)&=&0.59(10)\,,\\
w_0\Lambda^{\textrm{MiniMOM}}(\betal=1.9) &=& 0.70(9)\,,
\end{eqnarray}
are compatible within the errors for the two different bare gauge couplings we have analyzed. 

The knowledge of the $\Lambda$-parameter in the $\overline{\textrm{MS}}$-scheme is a crucial starting ingredient for the renormalization of energy-momentum tensor and of the supercurrents of $\mathcal{N}=1$ SYM using the Wilson Flow following the approach of Ref.~\cite{SUZ1,SUZ2,SUZ3}. Using the result at $\betal=1.9$ and the conversion factors coming from the perturbative expansion of the ghost-ghost-gluon vertex of Ref.~\cite{SME1,CHE}, we can also compute the $\Lambda$-parameter in the $\overline{\textrm{MS}}$-scheme, which is
\begin{equation}
w_0\Lambda^{\overline{\textrm{MS}}}(\betal=1.9)=0.385(50)\,.
\end{equation}
In QCD units, we obtain $\Lambda^{\overline{\textrm{MS}}} = 336(44)$ MeV, a value quite similar to the recent determinations of $\Lambda^{\overline{\textrm{MS}}}$ in full QCD \cite{FLAG}. 

\section{$N_f=2$ adjoint QCD}
\label{sec:nf2res}
Our main ensembles for $N_f=2$ AdjQCD have been generated at two different bare lattice gauge couplings $\betal=1.5$ and $\betal=1.7$. An additional set of simulations at $\betal=2.2$ is also analyzed, but these  ensembles should be considered with special care since the Polyakov loop in spacial directions shows already indications for a transition to the deconfined phase at these parameters. The detailed analysis about the simulations, the properties of the bound spectrum and the scaling of the mode number of the Dirac operator has been published in Ref.~\cite{BER2}.

\begin{figure}
\centering
 \subfigure[$\betal=1.5$ and $V=32^3\times 64$]{\includegraphics[width=.47\textwidth]{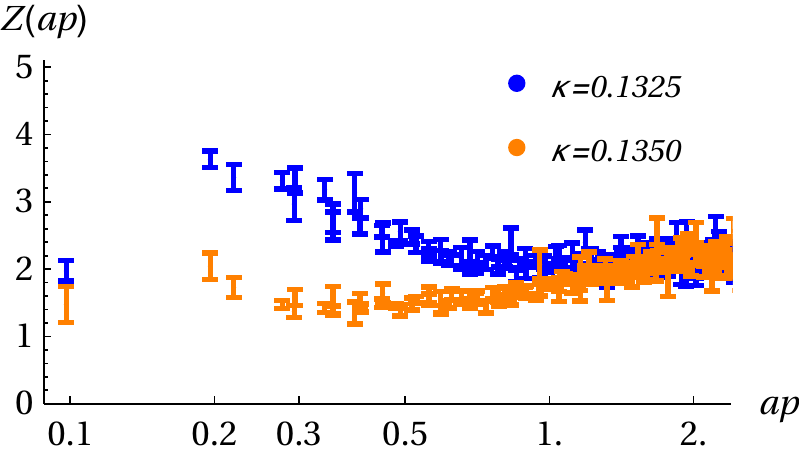}\label{nf2gluon150}}
 \subfigure[$\betal=1.5$ and $V=32^3\times 64$]{\includegraphics[width=.47\textwidth]{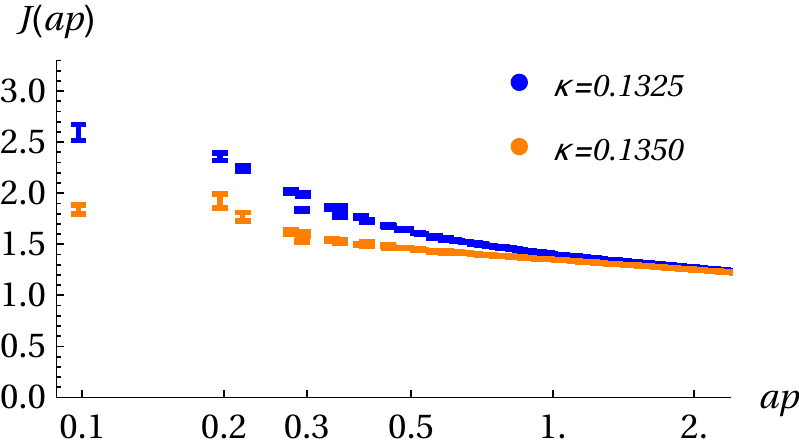}}
 \subfigure[$\betal=1.7$ and $V=32^3\times 64$]{\includegraphics[width=.47\textwidth]{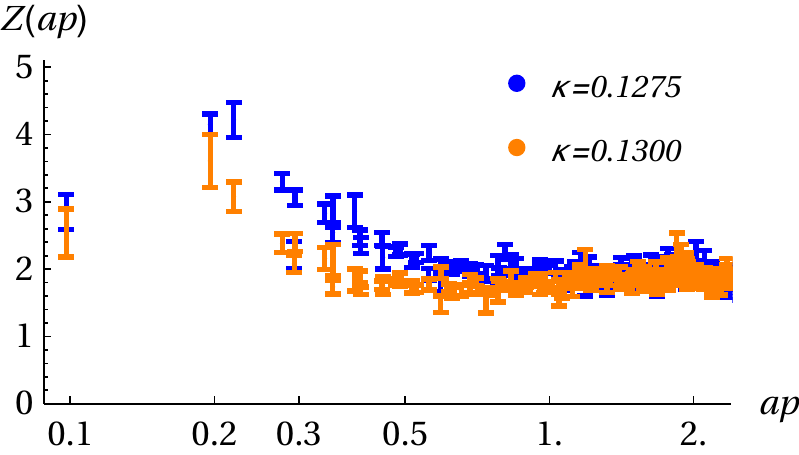}}
 \subfigure[$\betal=1.7$ and $V=32^3\times 64$]{\includegraphics[width=.47\textwidth]{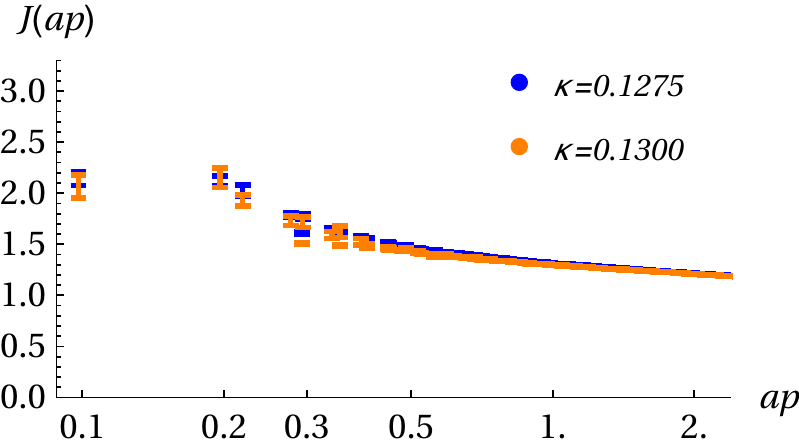}}
 \caption{Bare gluon and ghost dressing functions for $N_f=2$ AdjQCD as a function of the lattice momentum $ap$ at $\betal = 1.5$ and $\betal=1.7$.}\label{nf2gluonghost}
\end{figure}
The gluon and the ghost dressing functions have clearly a flatter behavior than in $\mathcal{N}=1$ SYM, especially at large momentum, see Fig.~\ref{nf2gluonghost}. In particular, the gluon dressing function develops a plateau as the fermion mass is decreased. At $\betal=1.5$ there is even a critical mass at which the gluon dressing function turns from a decreasing to an increasing function of the momenta, see Fig.~\ref{nf2gluon150}. The ghost dressing function has instead always a decreasing behavior. The physical consequence is that gluon and ghost propagators are becoming both divergent in the infrared limit as the fermion mass is decreased, at least up to the region of momenta explored in our simulations. This is consistent with the DSE data for a conformal theory \cite{HOP1,HOP2}.

\begin{figure}[t]
\centering
 \subfigure[$\betal=1.5$ and $V=32^3\times 64$]{\includegraphics[width=.47\textwidth]{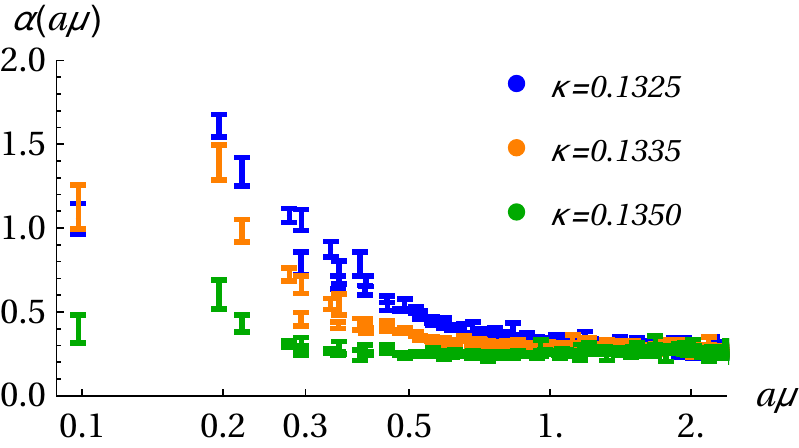}\label{nf2beta150}}
 \subfigure[$\betal=1.7$ and $V=32^3\times 64$]{\includegraphics[width=.47\textwidth]{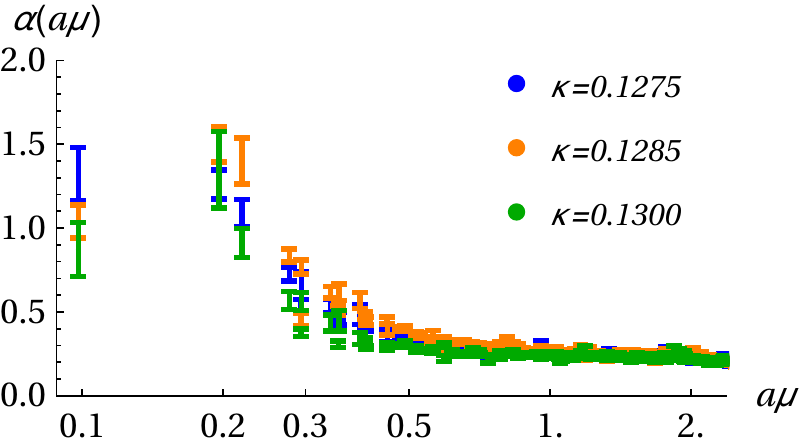}\label{nf2beta170}}
 \subfigure[$\betal=2.25$ and $V=32^3\times 64$]{\includegraphics[width=.47\textwidth]{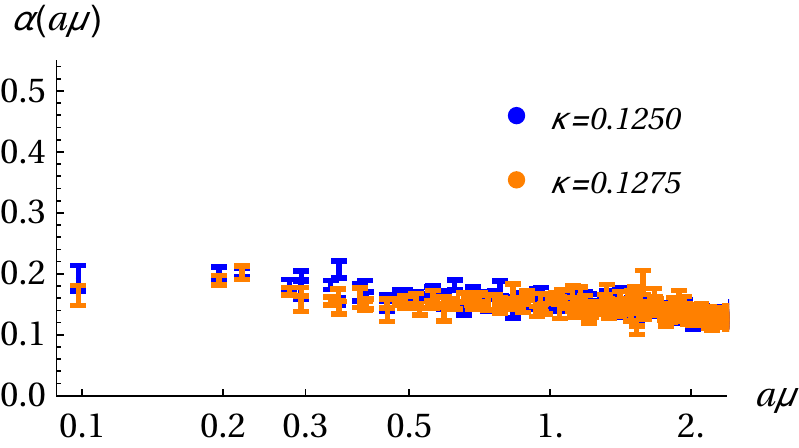}\label{nf2beta225}}
 \subfigure[$\betal=1.7$]{\includegraphics[width=.47\textwidth]{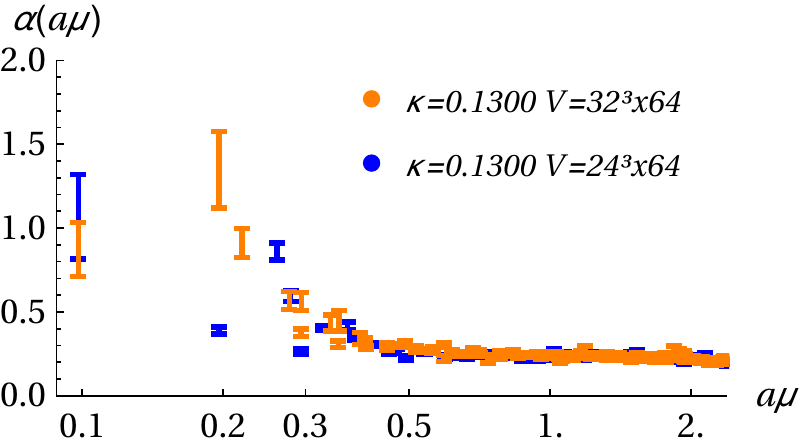}\label{nf2beta170finitevolume}}
 \caption{Strong coupling constant for $N_f=2$ AdjQCD as a function of the renormalization scale $a\mu$ at different $\betal$. The results show a clear fermion mass dependence and the strong coupling effectively freezes  toward the massless limit.}\label{nf2beta}
\end{figure}

The running coupling is presented for our three different bare lattice couplings in Fig.~\ref{nf2beta}. In contrast to $\mathcal{N}=1$ SYM, we observe a clear mass dependence of the running coupling at $\betal=1.5$ and $\betal=1.7$. At the smallest fermion mass we have simulated ($\kappa=0.1350$ and $\betal=1.5$), the running coupling develops a large plateau, which is the result of a cancellation between the upward tendency of the gluon dressing function against the decreasing behavior of the ghost dressing function. The plateau extends to the full scaling region at our largest $\betal=2.25$, where $\alpha(\mu)$ has a very weak dependence on $\mu$. 

Besides this plateau, there is a considerable running of coupling the far infrared region  at $\beta=1.5$ and $1.7$, which is decreased for smaller fermion masses. This far infrared part is also subject to considerable finite volume effects at scales smaller than $a\mu \sim 0.4$, see Fig.~\ref{nf2beta170finitevolume}. 

We don't see a significant running in the ultraviolet part. Due to the flat behavior of the strong coupling, the fit to perturbation theory is very unstable and it is impossible to estimate the ultraviolet scale $\Lambda$.

The general behavior with a nearly zero running of the coupling over a large energy scale is consistent with the expectations for an infrared fixed point. The fermion mass is a relevant direction at the fixed point and is responsible for the running in a region that moves further to the infrared the smaller the fermion mass is. Only the gluon dressing function at the smallest fermion masses of $\beta=1.5$ shows a decreasing behavior in the infrared as expected from the scaling form \eqref{eq:scaling} with a positive $\rho$. From a fit we obtain a value of around $\rho\sim 0.08$ which is considerably smaller than the value obtained in the DSEs approach for fundamental fermions. However, all of our other results are also quite consistent with the alternative scenario where both, ghost and gluon dressing functions become almost constant in the plateau region of the effective coupling.

Unfortunately we are not able to connect the running to the perturbative one since up to the highest energies we are able to explore the running seems to be dominated by the influence of the fixed point. There are two explanations, either our energy scales are still too low, or we are at couplings above the fixed point value and hence in a region not connected to the ultraviolet fixed point. 

\begin{figure}
\centering
 \subfigure[$\betal=1.5$ and $V=32^3\times 64$]{\includegraphics[width=.49\textwidth]{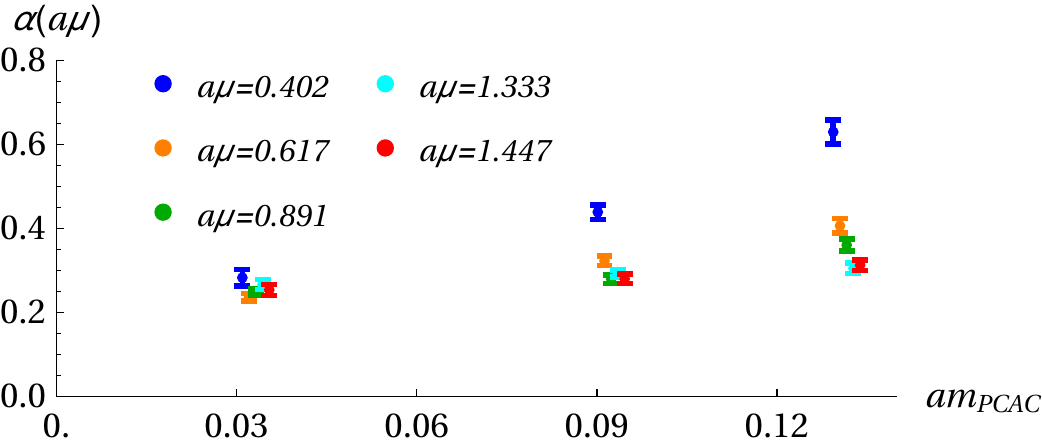}\label{nf2massdependence150}}
 \subfigure[$\betal=1.7$ and $V=32^3\times 64$]{\includegraphics[width=.49\textwidth]{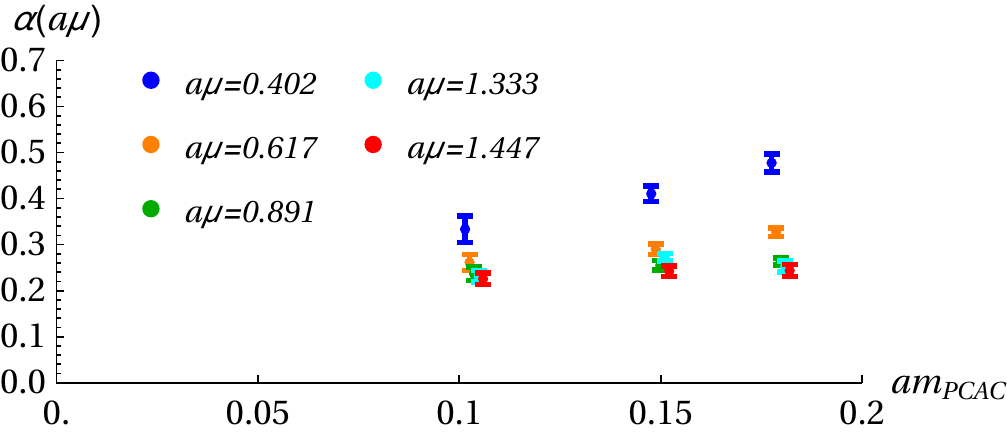}}
 \caption{Mass dependence of the running coupling for $N_f=2$ AdjQCD as a function of the bare $a m_{\textrm{PCAC}}$ at different $a\mu$. The data shown come from the ensembles at $\betal=1.5$ and $\betal=1.7$, (there is a tiny horizontal offset for each $\mu$ to make the individual points visible). All the couplings collapse to unique point at $a m_{\textrm{PCAC}} \simeq 0.03$, corresponding to $\kappa=0.1350$ (green points of Fig.~\ref{nf2beta150}). At the smallest $\kappa=0.1325$, the red and cyan points are still compatible within the errors ($\alpha^{\kappa=0.1325}(1.333) = 0.305(13)$ and $\alpha^{\kappa=0.1325}(1.447) = 0.313(12)$), meaning that the (near-)zero of the $\betal$-function can survive even for non-zero fermion masses.}\label{nf2massdependence}
\end{figure}
It is worth to investigate in more detail the dependence of the running coupling on the fermion mass and the deviations form $\als$ being a constant.  The coupling reaches its value at zero mass up to a correction that depends on the scale $\mu$ as can be clearly seen in Fig.~\ref{nf2massdependence}. In Fig.~\ref{nf2massdependence150}, couplings at different scales collapse to a unique point up to our statistical accuracy at $a m_{\textrm{PCAC}} \simeq 0.03$, and even at fermion masses as large as $m_{\textrm{PCAC}}=0.14$, many $\als(\mu)$ are still compatible within the errors for sufficient large $\mu$.  Hence up to our statistical accuracy the results are consistent with a zero of the $\beta$-function in the infrared region at small enough fermion masses. These observations support the evidences from previous lattice investigations that the $N_f=2$ AdjQCD theory is infrared conformal \cite{DEG,RAN}.

The fact that the fermion mass affects rather the far infrared part is quite expected and can be understood by comparison with perturbative QCD. The calculation of the first two coefficients of the $\beta$-function in the MOM scheme as a function of the quark mass is known from Ref.~\cite{YOS}. The main result is that $b_0$ and $b_1$ converge quite rapidly for $\mu^2 > m_f^2$ to their massless limit. Therefore, for small enough fermion masses and higher energy scales, we expect the running of $\als$ to be as slow as the running at zero fermion mass. A further study of massive schemes for near-conformal theories has been presented in Ref.~\cite{DIE1} in the context of NSZV-inspired $\beta$-functions.

There are two possible interpretations of the form of the fermion mass dependence, which are both consistent with our data. The first one is to assume that there is a real and exact zero of the beta function already at a finite fermion mass. Our simulations at $\betal=1.5$ have been done at a rather strong coupling, since this value of $\betal$ is just above the bulk transition in our lattice setup. In the infrared conformal scenario, the strong coupling region above the infrared fixed point is not connected to ultraviolet fixed point. Hence weak couplings below the IR fixed point are not reached when starting from the strong coupling side. Since the coupling has to run towards the lower value at the infrared fixed point, depending on the scheme, a backward running of the coupling is expected. As shown in \cite{Giedt:2011kz}, this might arise from extrapolation of a forward running at larger fermion masses towards the massless limit. This implies that at a finite mass the extrapolations should intersect implying a zero running of the coupling. The general form of our results are indeed consistent with the one obtained from a step-scaling methods in Ref.~\cite{Giedt:2011kz}, but as pointed out in this reference the uncertainties of the extrapolations do not allow a discrimination of the backward running from a zero running in the massless limit.

\begin{figure}
\centering
 \subfigure[$\betal=1.5$ and $V=32^3\times 64$]{\includegraphics[width=.47\textwidth]{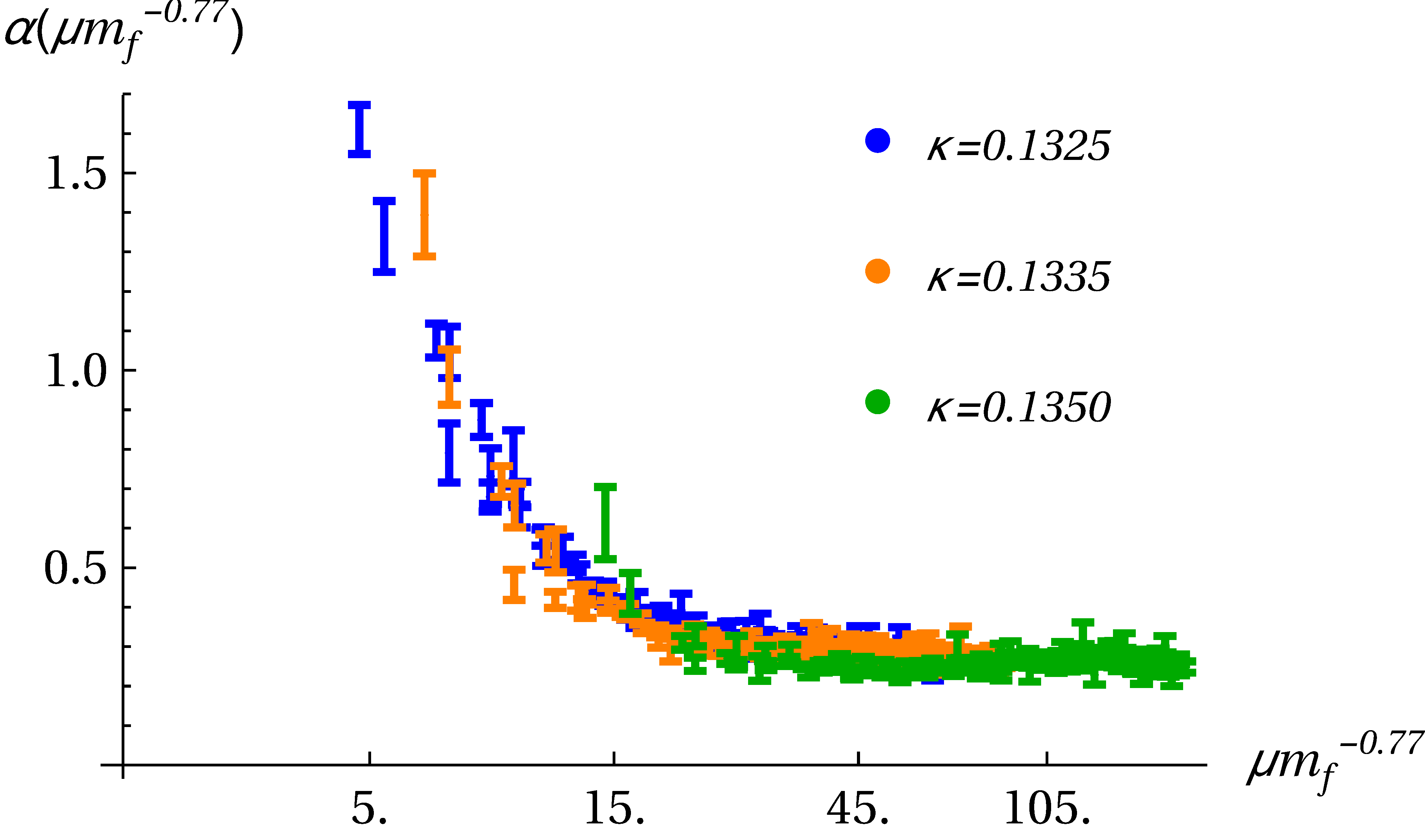}}
 \subfigure[$\betal=1.7$ and $V=32^3\times 64$]{\includegraphics[width=.47\textwidth]{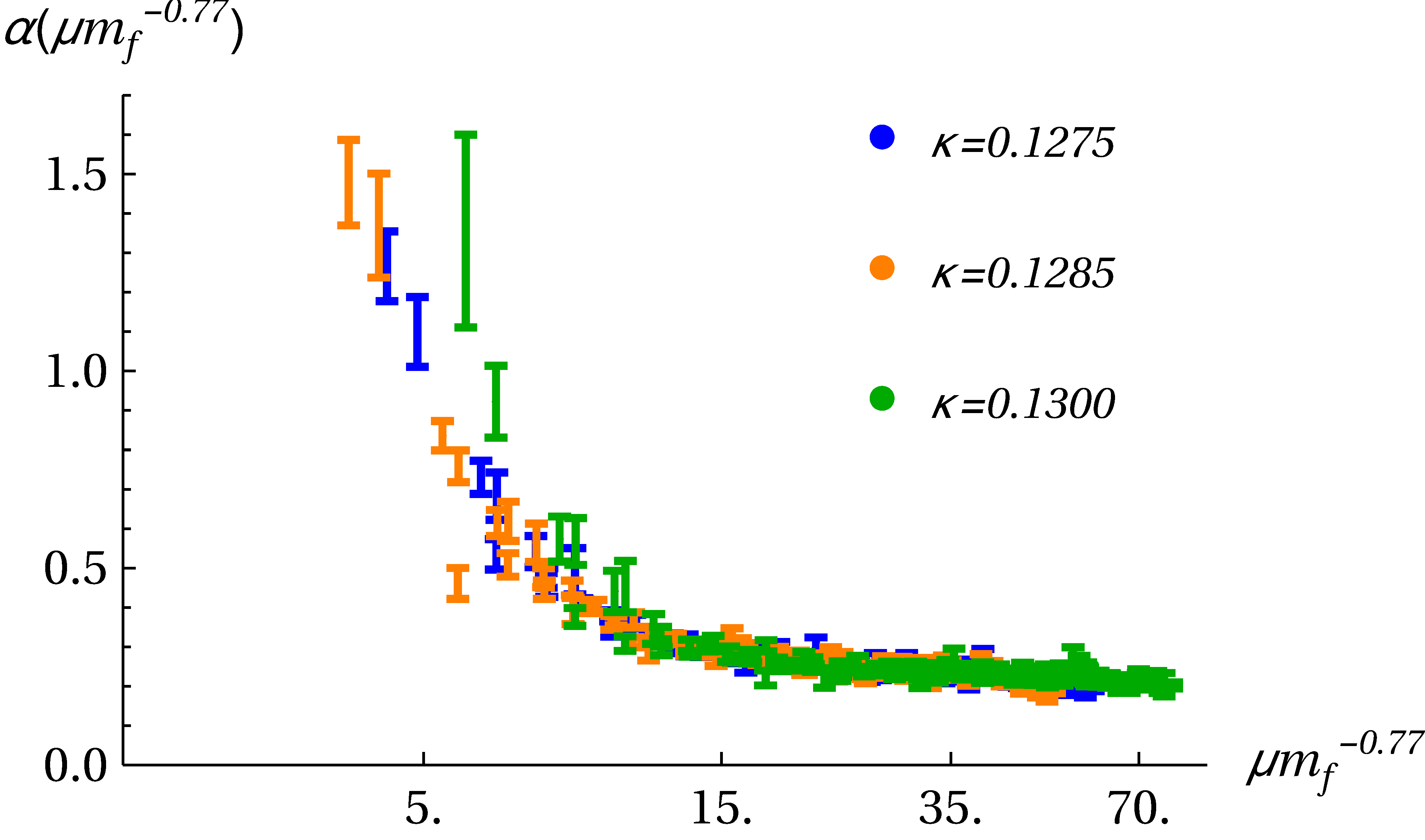}}
 \caption{``Hyperscaling'' of the running coupling for $N_f=2$ AdjQCD at $\betal = 1.5$ and $\betal=1.7$ as a function of $\mu \left(m_{\textrm{PCAC}}\right)^{-1/(1+\gamma^*)} \approx \mu \left(m_{\textrm{PCAC}}\right)^{-0.77}$. The strong coupling at different fermion masses appears to collapse toward a universal function. Some deviation is observed for small $\mu$ at the smallest fermion mass, probably due to finite volume effects.}\label{nf2hyperscaling}
\end{figure}

The alternative interpretation is that the running goes to zero only in the massless limit as dictated by hyperscaling. An infrared conformal theory does not possess any dimensionful scale related to long-range physics. In the chiral limit, all masses of bound states, including decay constants and condensates, extrapolate to zero according to a universal scaling
\begin{equation}\label{gamma_scaling}
 M \propto m^{\frac{1}{1+\gamma^*}}\,,
\end{equation}
where $\gamma^*$ is the mass anomalous dimension of the theory at the infrared fixed point \cite{DEB2}. Our previous investigations confirmed the validity of the scaling relation of the mass spectrum for the same ensembles, see Ref.~\cite{BER2}. As a consequence, the momenta cannot be converted to physical units by choosing a common scale for all fermion masses in a confining theory. However, the almost linear behavior of $\alpha(\mu)$ in Fig.~\ref{nf2massdependence} suggests that a possible universal function can be found for large enough momenta if the lattice scale $a \mu$ is measured in units of the fermion mass $m_{\textrm{PCAC}}$ as
\begin{equation}
\label{eq:hyperscaling}
 a\mu \rightarrow \frac{\mu}{(m_{\textrm{PCAC}})^{\frac{1}{1+\gamma^*}}}\,,
\end{equation}
with $\gamma^*$ equal to $\gamma^* \approx 0.3$ \cite{BER2}. This rescaling is equivalent to a conversion of the lattice spacing $a$ in units of the fermion mass with its appropriate dimension, employing a ``mass-dependent'' scale-setting scheme. The result of this momentum rescaling is presented in Fig.~\ref{nf2hyperscaling}. All points collapse to a unique curve except those at very small momenta, affected by finite volume effects and by the typical peak structure of $\als$ in momentum schemes. The scaling observed for the running coupling as a function of the fermion mass is also in agreement with the hypothesis that $N_f=2$ AdjQCD is an infrared conformal theory: in the infrared regime the coupling  runs only for non-vanishing values of the fermion mass, which break the conformal behavior explicitly, in a form that must scale as \eqref{gamma_scaling} close to the infrared fixed point. The main conclusion of the investigations of MWT is that all our ensembles are in a regime that is dominated by the infrared fixed point with the mass as the only relevant direction.

\section{$N_f=3/2$ adjoint QCD}
\label{sec:nf32res}
\begin{figure}[t]
\centering
 \subfigure[$\betal=1.5$ and $V=24^3\times 48$]{\includegraphics[width=.47\textwidth]{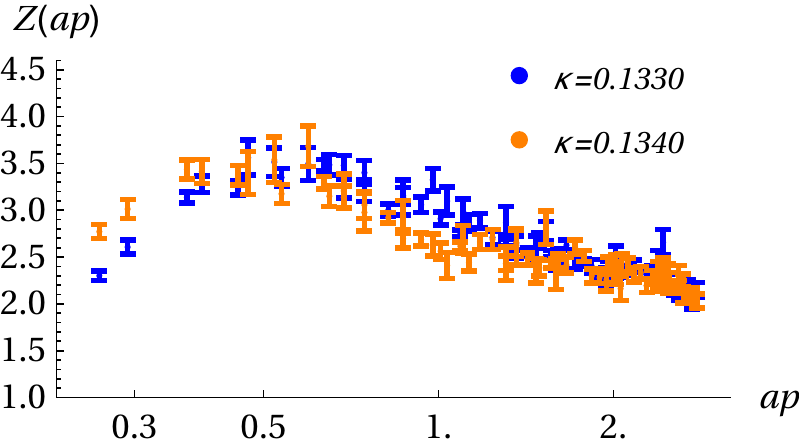}\label{nf32gluon150a}}
 \subfigure[$\betal=1.5$ and $V=24^3\times 48$]{\includegraphics[width=.47\textwidth]{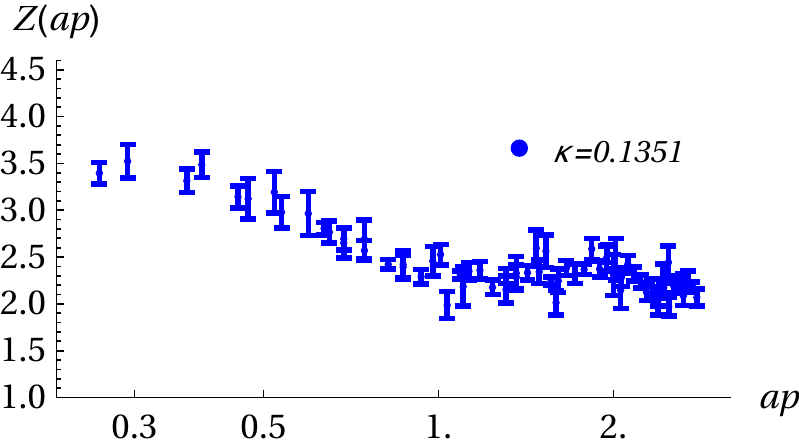}\label{nf32gluon150b}}
 \subfigure[$\betal=1.5$ and $V=24^3\times 48$]{\includegraphics[width=.47\textwidth]{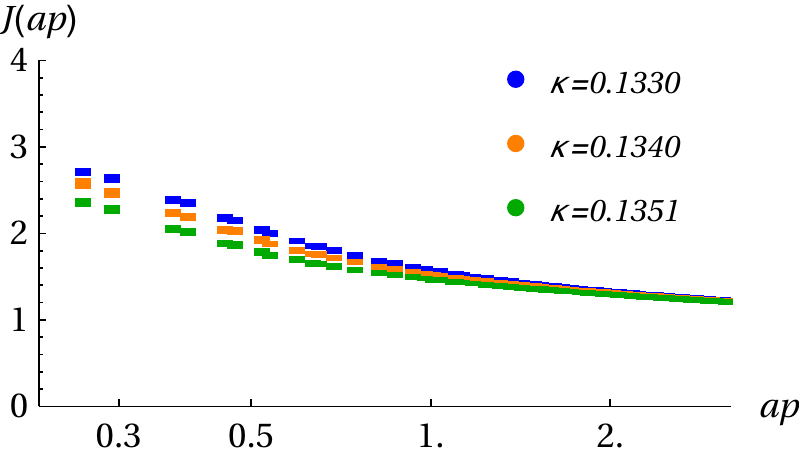}\label{nf32ghost150}}
 \caption{Bare gluon and ghost dressing functions for $N_f=3/2$ AdjQCD as a function of the lattice momentum $ap$ at $\betal = 1.5$.}
\end{figure}

We have generated ensembles at three different $\betal$ for $N_f=3/2$ AdjQCD. The target of our analysis is to check whether the running coupling is similar to $N_f=2$ AdjQCD or it is rather close to $\mathcal{N}=1$ SYM, in order to understand how the transition from a confining to a conformal theory occurs when the number of interacting fermions is increased.

For the two ensembles simulated with the largest $a m_{PCAC}$ at $\betal=1.5$ the gluon dressing function has a downward tendency in the deep infrared region similar to the one observed in pure gauge SU(2), see Fig.~\ref{nf32gluon150a}. The dressing function $Z(ap)$ becomes monotonous in the whole considered range of momenta only at the smallest fermion mass, see Fig.~\ref{nf32gluon150b}. The ghost dressing function $J(ap)$ shows instead a mild dependence both on $ap$ and on the fermion mass, see Fig.~\ref{nf32ghost150}.

\begin{figure}
\centering
 \subfigure[$\betal=1.5$ and $V=24^3\times 48$]{\includegraphics[width=.47\textwidth]{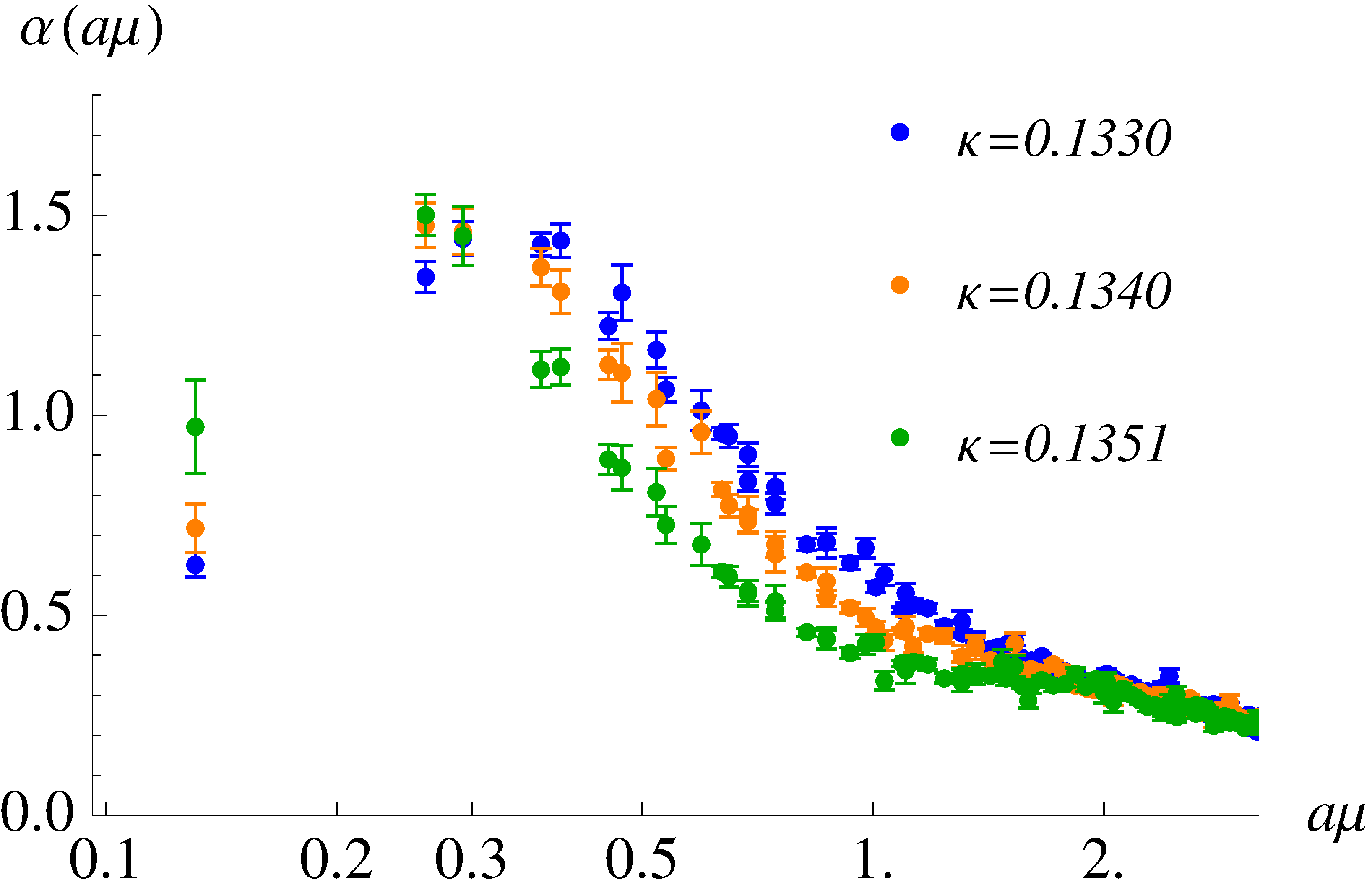}\label{nf321500b}}
 \subfigure[$\betal=1.6$ and $V=24^3\times 48$]{\includegraphics[width=.47\textwidth]{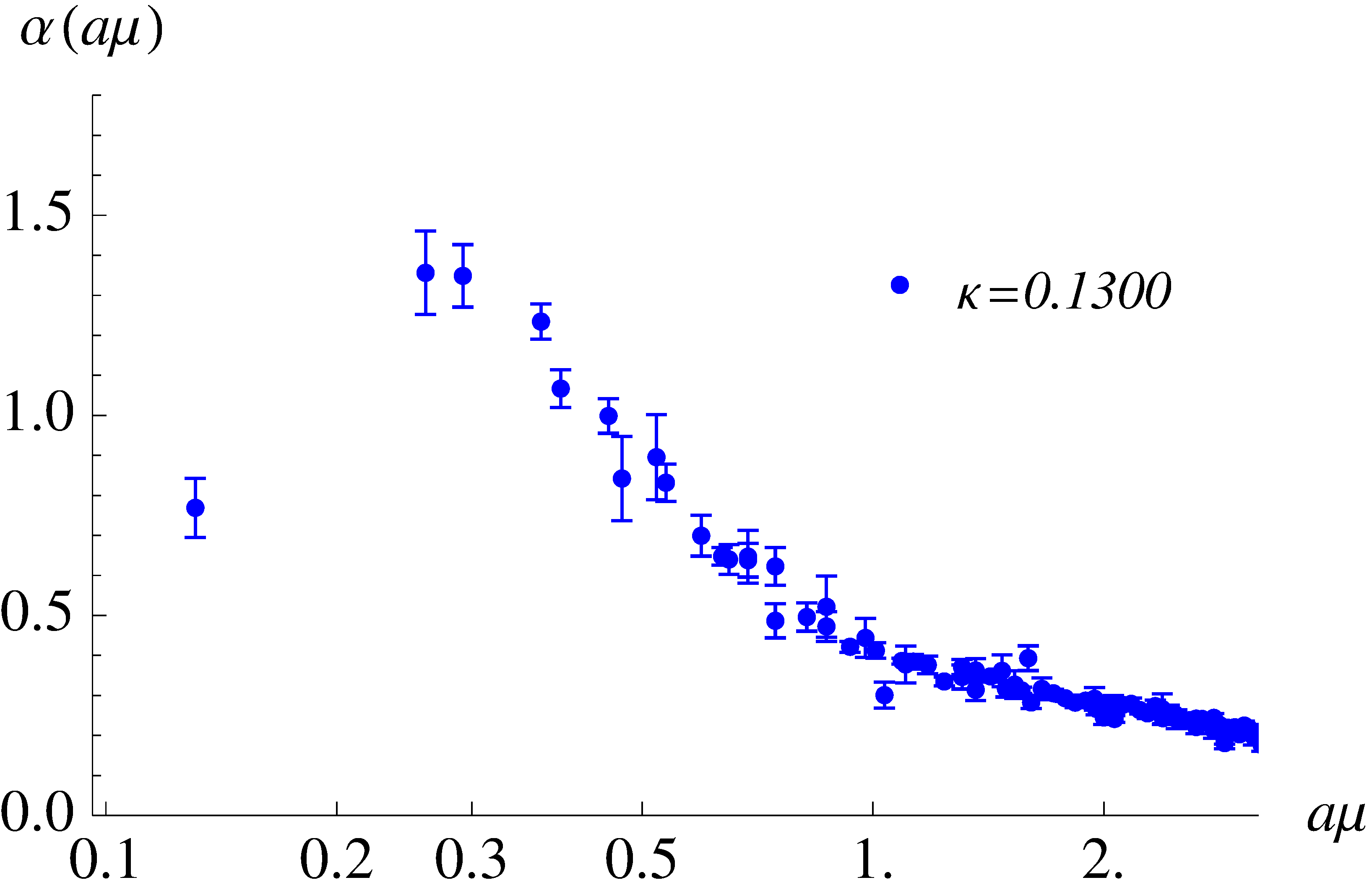}\label{nf321600b}}
 \subfigure[$\betal=1.7$ and $V=32^3\times 64$]{\includegraphics[width=.47\textwidth]{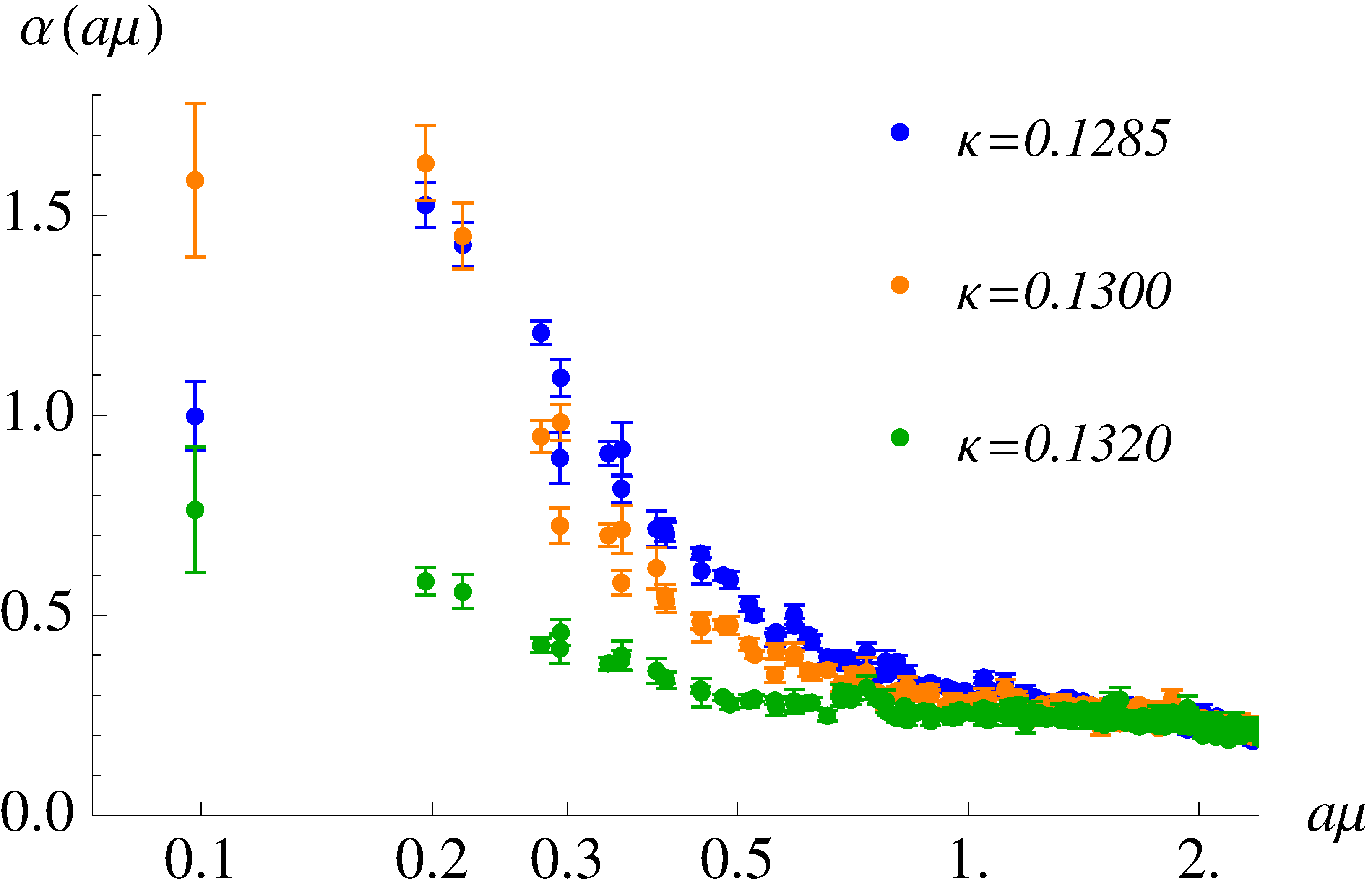}}
 \subfigure[$\betal=1.5$, $\kappa=0.1351$ and $V=24^3\times 48$]{\includegraphics[width=.47\textwidth]{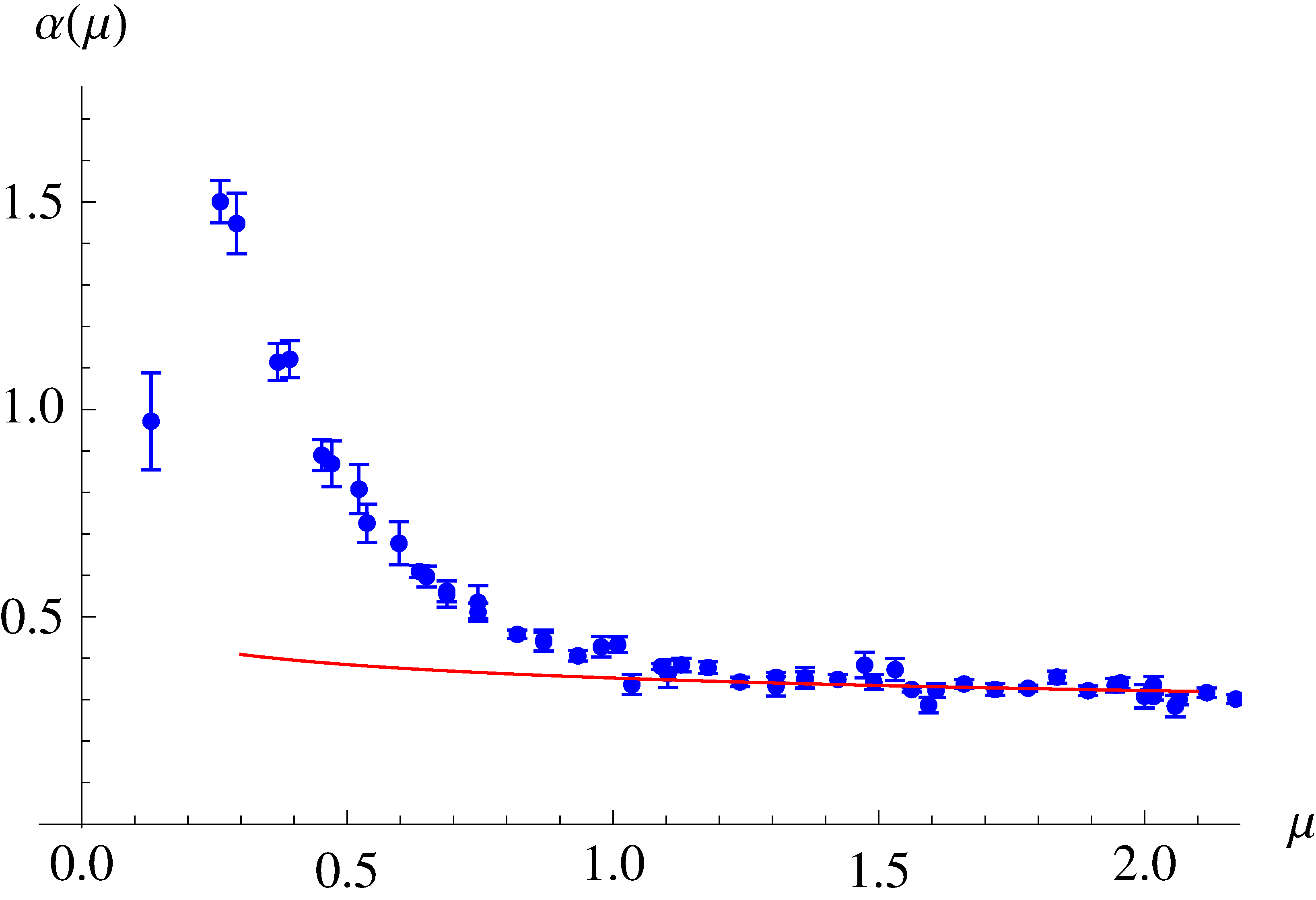}\label{nf321500bfitpt}}
 \subfigure[$\betal=1.6$, $\kappa=0.1300$ and $V=24^3\times 48$]{\includegraphics[width=.47\textwidth]{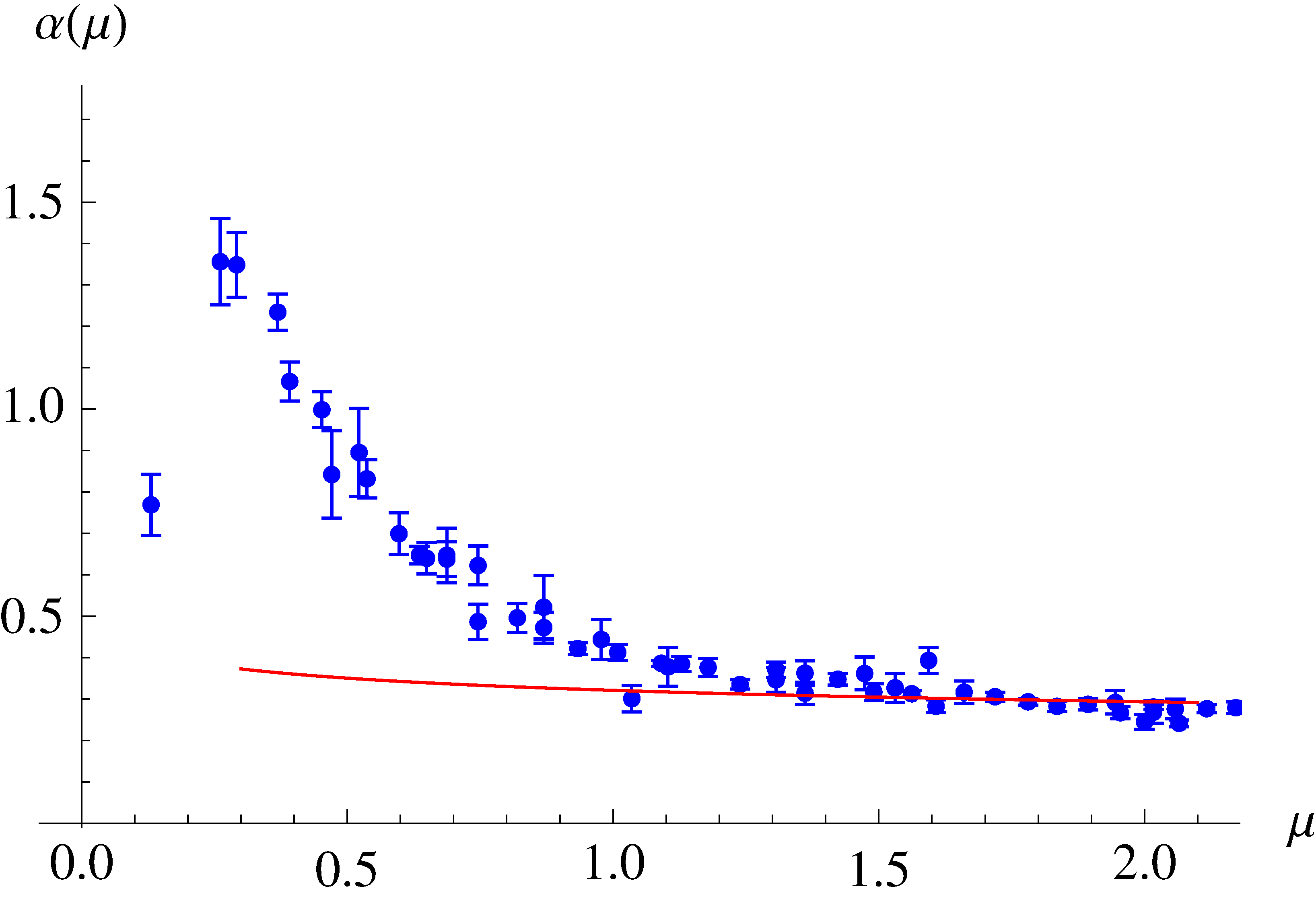}\label{nf321600bfitpt}}
 \subfigure[$\betal=1.7$, $\kappa=0.1320$ and $V=32^3\times 64$]{\includegraphics[width=.47\textwidth]{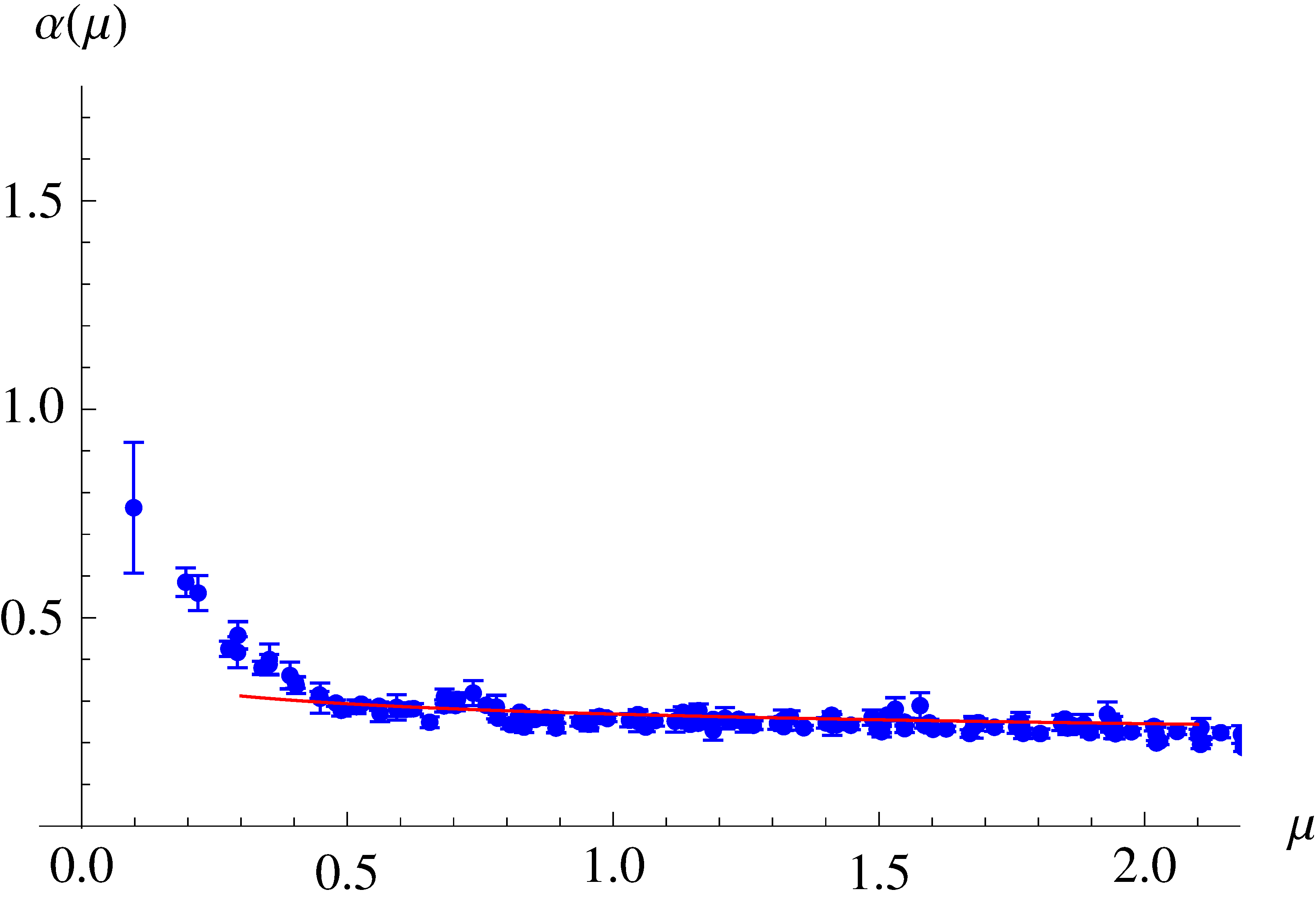}\label{nf321700bfitpt}}
 \caption{a-c) The running of the running coupling $\alpha(\mu)$ on a logarithmic scale $\mu$ for $N_f = 3/2$ AdjQCD. Many ensembles show that a rising tendency of $\alpha(\mu)$ that peaks in the infrared region. d-e) Fit of $\alpha(\mu)$ to perturbation theory at various $\betal$ with the largest $\kappa$ available plotted on a linear scale $\mu$. The fitting intervals are d) (1.3,2.05), e) (1.3,2.05), f) (0.9,2.15)}\label{nf32alpha}
\end{figure}

The running coupling has a downward tendency in the deep infrared region for our smallest $\mu$ (in particular at $\betal=1.5$ and $\betal=1.6$, see Fig.~\ref{nf321500b} and Fig.~\ref{nf321600b}), driven by the downward behavior of the gluon dressing function. As for $N_f=2$ AdjQCD, we observe a clear dependence of the running coupling on the fermion mass. The running of $\als$ is slow but still incompatible with zero, at least in the high momentum region where the effects of the non-vanishing fermion masses can be neglected.

In contrast to the $N_f=2$ case, the running coupling can be fitted to perturbation to check a possible relation to the continuum theory. The best fit of $\als(\mu)$ to perturbation theory with a reasonable $\chi^2$ and a wide range of $\mu$ comes from our data  at $\betal=1.7$ and $\kappa=0.1320$, due to the fact that the other ensembles have too large fermion masses that are not negligible even at quite large $\mu$ (see Fig.~\ref{nf321500bfitpt},  Fig.~\ref{nf321600bfitpt}, and Fig.~\ref{nf321700bfitpt}). Following the same procedure as for $\mathcal{N}=1$ SYM, we get
\begin{eqnarray}
a\Lambda^{\textrm{MiniMOM}}(\betal=1.5) & = & 0.0051(8)\,,\\
a\Lambda^{\textrm{MiniMOM}}(\betal=1.6) & = & 0.0055(4)\,,\\
a\Lambda^{\textrm{MiniMOM}}(\betal=1.7) & = & 0.0054(5)\,.
\end{eqnarray}
All quoted errors are purely systematic corresponding to variations coming from different choices of the fitting intervals. The value of $\Lambda$ in lattice units is significantly smaller than the masses of hadrons in lattice units at the same parameters, for instance the glueball $0^{++}$ mass $a m_{0^{++}} \simeq 0.25(3)$. The contrast to the well-known confining gauge theories is quite significant, in QCD the value of the intrinsic ultraviolet scale $\Lambda$ is comparable to the hadron scales in the various MOM schemes. This is in accordance with the observation that the fermion mass sets the scale of the mass spectrum of this theory approximately according to \eqref{eq:hyperscaling}. It is, furthermore, remarkable how weak the dependence of $a\Lambda$ on the bare lattice coupling is. Even though these observations are not enough to provide a strong evidence for a conformal behavior as in the $N_f=2$ case, they are clearly much different from QCD-like theories.

\section{$N_f=1$ adjoint QCD}
\label{sec:nf1res}
\begin{figure}
\centering
 \subfigure[$\betal=1.75$ and $V=24^3\times 48$]{\includegraphics[width=.47\textwidth]{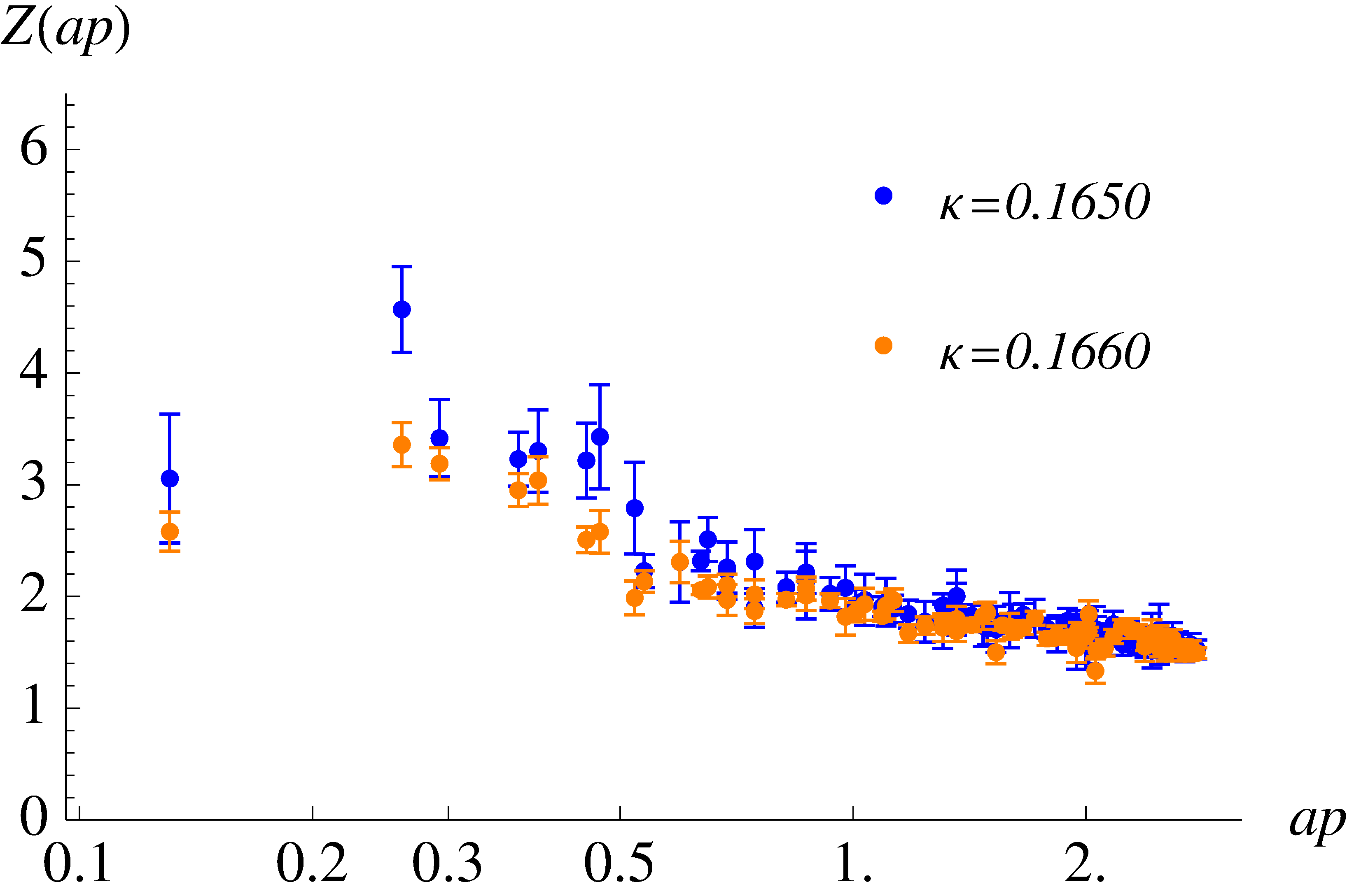}\label{nf22gluon175}}
 \subfigure[$\betal=1.75$ and $V=24^3\times 48$]{\includegraphics[width=.47\textwidth]{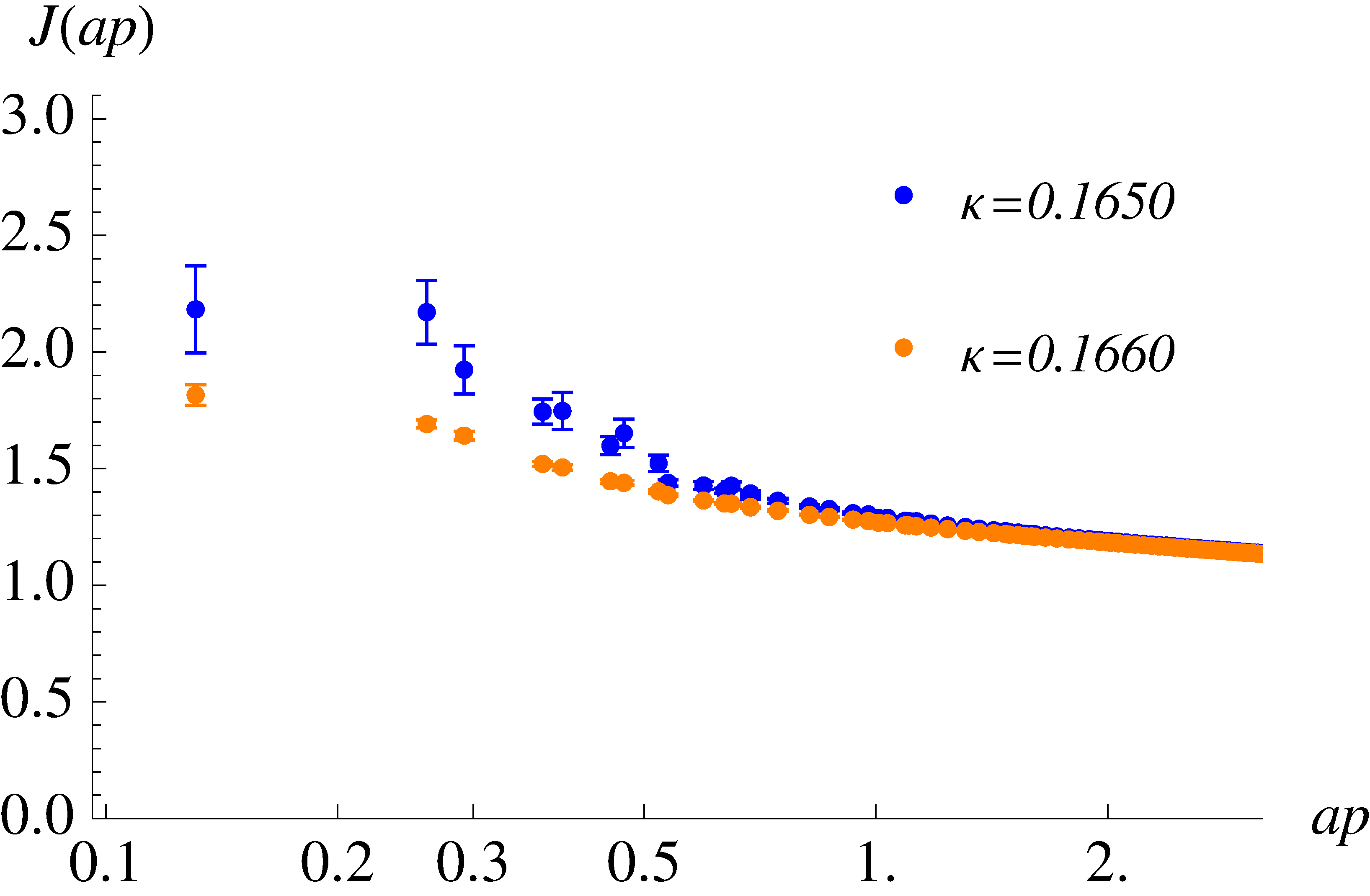}\label{nf22ghost175}}
 \caption{Bare gluon and ghost dressing functions for $N_f=1$ AdjQCD as a function of the lattice momentum $ap$ at $\betal = 1.75$.}
\end{figure}

Since moving from four to three Majorana fermions coupled to SU(2) gluodynamics has been sufficient for a significant change in the running of $\alpha$, it is interesting to check what happens to $N_f=1$ AdjQCD. We have generated a set of ensembles at $\betal=1.75$ and several values of $\kappa$. The largest statistics has been collected from two ensembles at $\kappa=0.1650$ and $\kappa=0.1660$ with a lattice size of $24^3\times 48$. The ghost and the gluon dressing functions have a similar tendency as observed for $N_f=3/2$ AdjQCD (see Fig.~\ref{nf22gluon175} and Fig.~\ref{nf22ghost175}), with a peak of $Z(ap)$ in the infrared and a mild dependence of the dressing functions on the fermion mass. 

\begin{figure}
\centering
 \subfigure[$\betal=1.75$ and $V=24^3\times 48$]{\includegraphics[width=.47\textwidth]{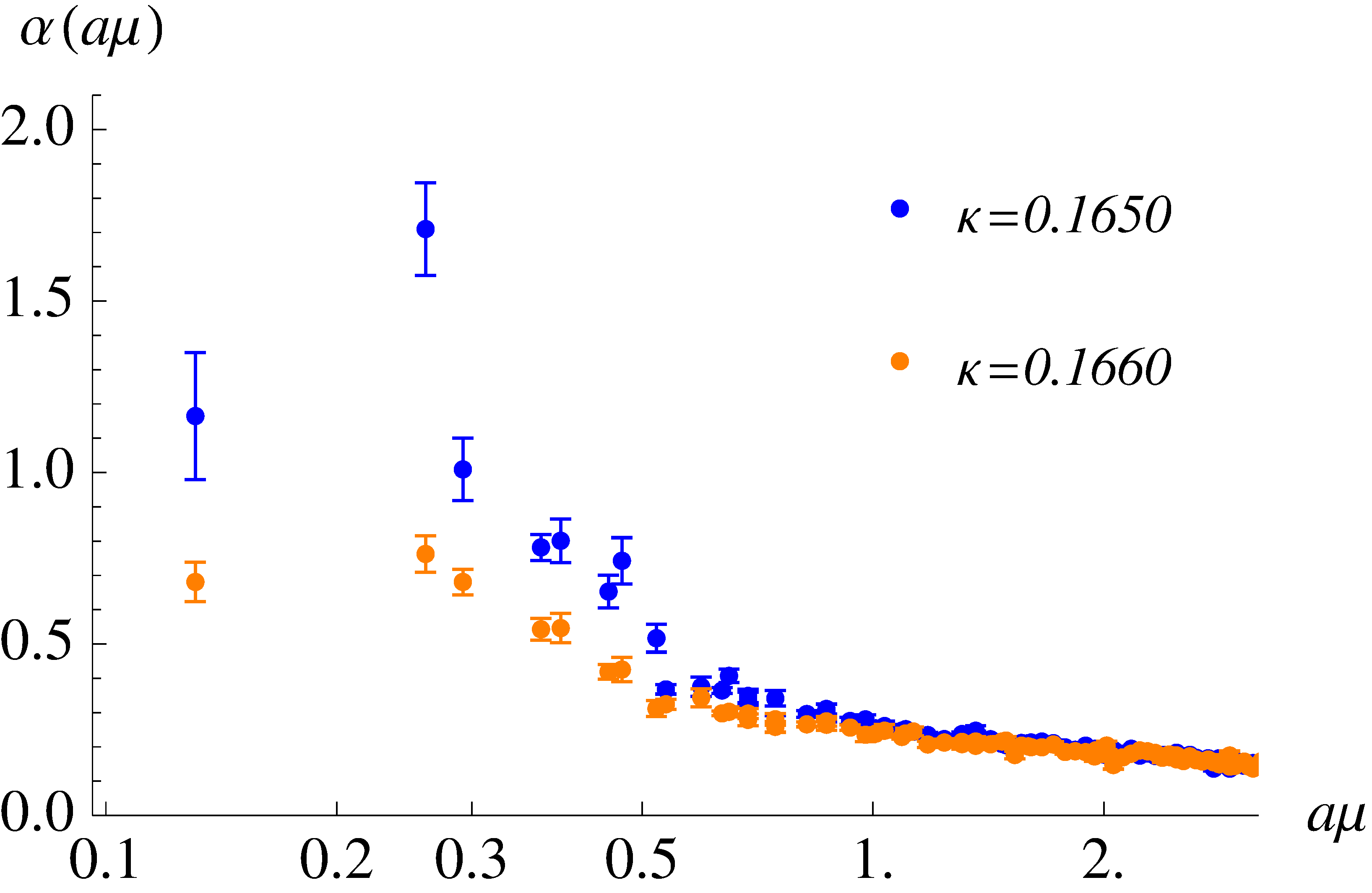}\label{nf221750b}}
  \subfigure[$\betal=1.75$ and $\kappa=0.165$]{\includegraphics[width=.47\textwidth]{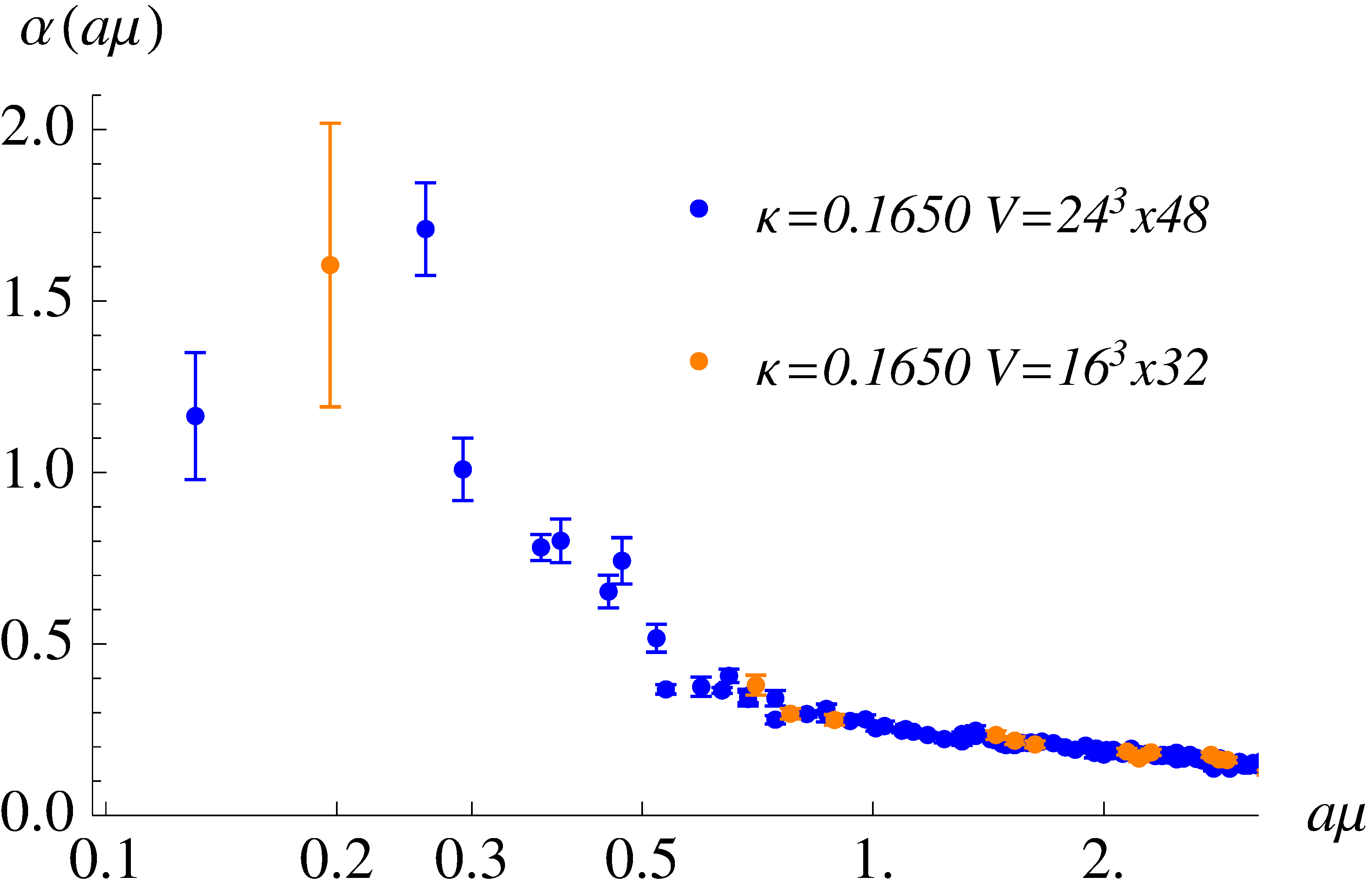}\label{nf221750bfv}}
 \caption{a) The running of the running coupling $\alpha(\mu)$ on a logarithmic scale $\mu$ for the $N_f = 1$ AdjQCD theory. b) Comparison of finite volume effects at $\kappa=0.1650$.}\label{nf22alpha}
\end{figure}

The running coupling is shown in Fig.~\ref{nf22alpha}. Finite volume effects seems to be under control, see Fig.~\ref{nf221750bfv}. There is a dependence of the strong coupling on the fermion mass at small $\mu$, while at larger energy scales the running approaches well the predicted perturbative $\beta$-function of the massless theory.

\begin{figure}
\centering
 \includegraphics[width=.67\textwidth]{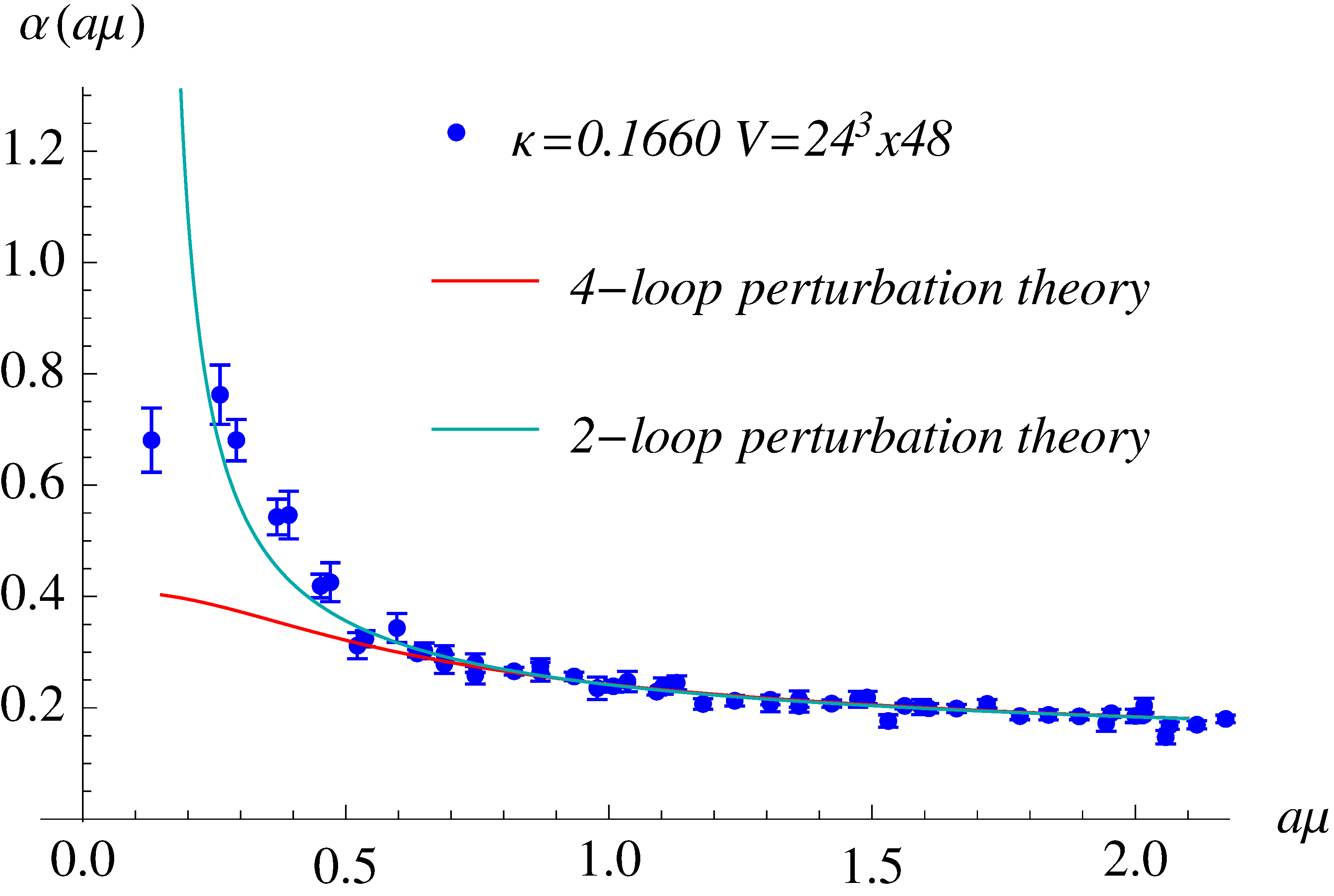}
 \caption{Comparison of the fit of $\als(\mu)$ to two and four loops perturbation theory plotted on a linear scale $\mu$. The fitting interval is (0.47,2.2). The fit to two-loop perturbation theory might appear better, but the agreement at low $a\mu$ is most likely accidental, given that for $a\mu < 0.3$ the effects of the fermion mass on $\alpha(\mu)$ are not negligible, see Fig.~\ref{nf221750b}.}\label{nf22alphapt}
\end{figure}

It is therefore interesting to compare the fits of $\als$ to two and four loop perturbation theory. As before, we fit for the four loop running the ultraviolet scale and absolute normalization of $\als$ to the data of the smallest fermion mass ($\kappa=0.1660$) in a region with $a\mu > 0.5$ in order to ensure negligible fermion mass effects. The scale $\Lambda$ in lattice units is
\begin{equation}
a \Lambda^{\textrm{MiniMOM}}(\betal=1.7) = 0.20(3)\,,
\end{equation}
the quoted error is again systematic. The $\Lambda$-parameter has developed a quite large value with respect to the pion mass compared to the case of $N_f=3/2$. As for $\mathcal{N}=1$ SYM, it is of the same order of magnitude as the mass gap. In addition, it is worth to note that also two-loop perturbation theory might fit well our data at sufficiently large $\mu$, see Fig.~\ref{nf22alphapt}. The main conclusion from these investigations is that the scale where the strong coupling of $N_f=1$ AdjQCD might eventually develop an infrared fixed point is very small and beyond the reach of our current volumes and fermion masses. Further simulations at large volumes and smaller fermion masses would be required to reach a full definitive conclusion about its existence.

\section{Conclusions}

We have presented a lattice study of the gluodynamics in the Landau gauge for AdjQCD and observed that the properties of gluon propagators depend crucially on the number of adjoint fermions. We have determined the running coupling in the MiniMOM scheme and investigated how it changes when going from the QCD-like case to a theory in the conformal window. 

For $N_f=1/2$, i.~e.\ $\mathcal{N}=1$ SYM, the gluon propagator has, similar to QCD, a non-trivial infrared behavior that does not show a significant dependence on the fermion mass. The gluon dressing function develops a stronger dependence on the fermion mass the more fermions are coupled to the gluons. The behavior of the ghost dressing functions appears to be less dependent on the number of fermions, although a stronger flattening is observed towards the conformal window. 

The running coupling computed from the gluon and ghost dressing functions provides a clear evidence for the asymptotic freedom of adjoint QCD with $N_f=1/2$ and $N_f=1$ fermions. For $N_f=1$ AdjQCD, there could only be evidence for an infrared fixed point at very small scales compared to the one that we can reliably explore with our simulations. A very slow running of $\als$ is observed for $N_f=3/2$ AdjQCD. Consequently this theory appears to be close to or inside the conformal window.

We have estimated the $\Lambda$-parameter in the MiniMOM and in the $\overline{MS}$ scheme for $\mathcal{N}=1$ SYM. This result is crucial for the renormalization of the supercurrents using the methods proposed in Ref.~\cite{SUZ1,SUZ2,SUZ3}. We have determined the $\Lambda$-parameter in the MiniMOM scheme also for $N_f=1$ and $N_f=3/2$ AdjQCD, while the $\beta$-function of $N_f=2$ AdjQCD is strongly affected by the non-vanishing value of the fermion mass at the considered range of the scale $\mu$. The determined value of $\Lambda$ for $N_f=3/2$ AdjQCD is several orders of magnitude smaller than the mass gap. 

The results for $N_f=2$ AdjQCD, i.~e.\ MWT, are completely different from the QCD case. Consistent with the expectations for a conformal theory, one observes a plateau region with an almost constant $\als(\mu)$ and a strong fermion mass dependence in the far infrared. At the considered energy range the fermion mass seems to be the only relevant parameter that determines the running of $\als(\mu)$. This is consistent with the fact that the mass is the only relevant direction at the fixed point. The specific form of the fermion mass dependence can be either interpreted as an indication for backward running at zero fermion mass or in terms of hyperscaling. 

At our smallest PCAC mass, the gluon dressing functions of $N_f=2$ AdjQCD shows a similar functional form as the one expected from the DSE approach. Within the range of momenta that we can reliably explore in our simulations, the gluon propagator diverges in the infrared regime, consistent with the disappearance of the plateau of the propagator determined from the DSEs. Hence the general picture appears to be consistent with the DSE results for a conformal theory with fermions in the fundamental representation.

The study of the running of $\als$ for an infrared conformal theory requires special considerations concerning the tuning of the lattice parameters, the continuum and chiral extrapolations, and the fit to pertubation theory. Our results for $N_f=2$ AdjQCD do not show a significant dependence on the bare gauge coupling $\betal$. However, the investigated range of bare couplings should be enlarged to understand how the ultraviolet fixed point is approached and to clearly separate possible unphysical regions of bare lattice parameters beyond the infrared fixed point, where a flat or even a backward running could be expected \cite{Giedt:2011kz}.

\section{Acknowledgements}

We thank Gunnar S. Bali, Jacques C.R. Bloch, Biagio Lucini, Holger Gies, and Meinulf G\"ockeler for interesting discussions. We thank Gernot M\"unster and Pietro Giudice for reading and commenting the first version of the manuscript. We thank Istvan Montvay for his support in performing the numerical simulations.

The authors gratefully acknowledge the Gauss Centre for Supercomputing e.~V.\ (GCS) for providing computing time for a GCS Large-Scale Project on the GCS share of the supercomputer JUQUEEN at J\"ulich Supercomputing Centre (JSC) and on the supercomputer SuperMUC at Leibniz Computing Centre (LRZ). GCS is the alliance of the three national supercomputing centres HLRS (Universität Stuttgart), JSC (Forschungszentrum J\"ulich), and LRZ (Bayerische Akademie der Wissenschaften), funded by the German Federal Ministry of Education and Research (BMBF) and the German State Ministries for Research of Baden-Württemberg (MWK), Bayern (StMWFK) and Nordrhein-Westfalen (MIWF). Further computing time has been provided by the iDataCool cluster of the Institute for Theoretical Physics at the University of Regensburg. SP acknowledges support from the Deutsche Forschungsgemeinschaft (DFG) Grant No. SFB/TRR 55 and GB by Grant No. BE 5942/2-1.

\end{document}